\documentclass[12pt]{article}
\usepackage{enumerate}
\usepackage{amsmath}
\usepackage{amssymb}
\usepackage{epic}
\usepackage{epic,eepic}
\usepackage{fancyheadings}
\usepackage[dvips]{graphicx}
\usepackage{graphics}
\usepackage{graphicx}
\usepackage{latexsym}
\usepackage{ulem}
\usepackage{xspace}
\usepackage{lscape}
\usepackage{yhmath}

\allowdisplaybreaks

\usepackage{delarray}

\def\IntKer2{R_{X}(K)}

\newtheorem{Proposition}{Proposition}

\newtheorem{Remark}{Remark}

\begin{document}



\centerline{}
\vskip 3mm

\noindent \begin{center}
{\Huge Skewing Methods for Variance-Stabilizing Local Linear Regression Estimation}
\end{center}
\vskip 3mm

\vskip 5mm
\noindent Kiheiji NISHIDA

\vskip 3mm

\noindent
Lecturer, General Education Center, Hyogo University of Health Sciences.\\
1-3-6, Minatojima, Chuo-ku, Kobe, Hyogo 650-8530, JAPAN.\\
E-mail: kiheiji.nishida@gmail.com

\vskip 2mm

\vskip 3mm

\noindent ABSTRACT

It is well-known that kernel regression estimators do not produce a constant estimator variance over a domain. To correct this problem, Nishida and Kanazawa (2015) proposed a variance-stabilizing (VS) local variable bandwidth for Local Linear (LL) regression estimator. In contrast, Choi and Hall (1998) proposed the skewing (SK) methods for a univariate LL estimator and constructed a convex combination of one LL estimator and two SK estimators that are symmetrically placed on both sides of the LL estimator (the convex combination (CC) estimator) to eliminate higher-order terms in its asymptotic bias. To obtain a CC estimator with a constant estimator variance without employing the VS local variable bandwidth, the weight in the convex combination must be determined locally to produce a constant estimator variance. In this study, we compare the performances of two VS methods for a CC estimator and find cases in which the weighting method can superior to the VS bandwidth method in terms of the degree of variance stabilization.
\noindent
\\
\vskip 4mm



\vspace{-15mm}
\section{Introduction} \label{Introduction}

Suppose that $X_{1}, X_{2}, ..., X_{n}$ are i.i.d. explanatory random variables with density function $f_{X_{i}}(x_{i})$ on bounded support $I \in R$. The response $Y_{i}, i=1,2,..,n,$ is written as $Y_{i} = m(X_{i}) + U_{i}|X_{i}$, where $m(\cdot)$ is an $R \to R$ function of the $X_{i}$ and $U_{i}|X_{i}$'s, $i=1,...,n$ are disturbances independent with respect to $i$, and assumed to be independent of $X_{j}$, $i \neq j$, with the conditions $E_{U_{i}|X_{i}}[U_{i}|X_{i}=x_{i}] = 0$ and $E_{U_{i}|X_{i}}[U_{i}^{2}|X_{i}=x_{i}] = \sigma^{2}(x_{i})$.  Our aim is to estimate the regression mean, $m(x) = E(Y_{i}|X_{i}=x)$. 

Locally liner (LL) regression estimation is a refined method for achieving this aim (Fan and Gijbels, 1992). LL estimation involves estimating a straight line that is tangent to $m(x)$ and then plotting the intercept of the estimated straight line at every tangent point $x$, defined by $y(u) = \beta_{0} + \beta_{1}(u - x)$ through data pairs $(X_{i}, Y_{i})$, to choose $\beta_{0}$ and $\beta_{1}$ that minimize the weighted residual sum of squares,
\begin{eqnarray} \label{def_LL}
WRSS(\beta_{0}, \beta_{1}) = \sum_{i=1}^{n} \left[ Y_{i} - \beta_{0} - \beta_{1}(X_{i} - x) \right]^{2} K_{X} \left( \frac{X_{i} - x}{h} \right), \nonumber
\end{eqnarray}
where $K_{X}(\cdot)$ is a nonnegative and symmetric kernel function with adequate smoothness and $h=h(n)$ is a bandwidth as a function of the sample size $n$ satisfying $h \to 0$ and $nh \to \infty$ as $n \to \infty$. The minimizing pair $(\widehat{\beta_{0}}(x), \widehat{\beta_{1}}(x))$ is given by
\begin{eqnarray} \label{a_hat_b_hat}
\widehat{\beta_{0}}(x) = \frac{r_{0}(x)s_{2}(x) - r_{1}(x)s_{1}(x)}{r_{0}(x)s_{2}(x) - s_{1}^{2}(x)},\ \ \  \widehat{\beta_{1}}(x) = \frac{r_{1}(x)s_{0}(x) - r_{0}(x)s_{1}(x)}{s_{0}(x)s_{2}(x) - s_{1}^{2}(x)}, \nonumber
\end{eqnarray}
where
\begin{eqnarray} \label{r_i_s_i}
r_{j}(x) = \sum_{i=1}^{n} (X_{i} - x)^{j} K_{X} \left( \frac{X_{i}-x}{h} \right)Y_{i}, \ \ \ s_{j}(x) = \sum_{i=1}^{n} (X_{i} - x)^{j} K_{X} \left( \frac{X_{i}-x}{h} \right), \ \ \ j=0,1,2.\ \ \nonumber
\end{eqnarray}

Choi and Hall (1998) proposed a skewing (SK) method for LL estimation in the context of bias reduction. Skewing involves calculating the estimator at an off-center point $x_{0}$ slightly to the left or right of $x$, but nevertheless evaluating the estimator at $x$. An SK estimator estimates a straight line tangent to $m(x)$ at $x=x_{0}$ and is expressed by
\begin{eqnarray} \label{u_given_x}
\widehat{m_{h}}(x|x_{0}) &=& \widehat{\beta_{0}}(x_{0}) + \widehat{\beta_{1}}(x_{0})(x-x_{0}) \nonumber \\
&=& \frac{r_{0}(x_{0})s_{2}(x_{0})-r_{1}(x_{0})s_{1}(x_{0}) + \{ r_{1}(x_{0})s_{0}(x_{0})-r_{0}(x_{0})s_{1}(x_{0}) \}(x-x_{0})}{s_{0}(x_{0})s_{2}(x_{0})-s_{1}^{2}(x_{0})}. \nonumber
\end{eqnarray}
According to Choi and Hall (1998), if we denote the interval $x-x_{0}$ by $lh$, then the asymptotic bias of the SK estimator at $x$ with its tangent point being $x_{0}=x+lh$ is
\begin{eqnarray} \label{expansion}
\lefteqn{E_{X_{i}, Y_{i}} \{ \widehat{m_{h}}(x|x+lh) \left| X_{1}, X_{2}, ..., X_{n} \right. \}- m(x)} \nonumber \\
&=& \frac{1}{2}(\kappa_{2}-l^{2})m^{(2)}(x)h^{2} \nonumber \\
& & + \frac{l}{2} \left[ \frac{f^{(1)}(x)(\kappa_{2}^{2} - \kappa_{4})}{f_{X}(x)\kappa_{2}} m^{(2)}(x) + \left( \kappa_{2} - \frac{\kappa_{4}}{3 \kappa_{2}} - \frac{2l^{2}}{3} \right) m^{(3)}(x) \right]h^{3} \nonumber \\
& & + \frac{1}{2} \Biggl[ \Biggl( \left\{ \frac{f^{(2)}(x)}{2f_{X}(x)} - \left(\frac{f^{(1)}(x)}{f_{X}(x)}\right)^{2} \right\}(\kappa_{4} - \kappa_{2}^{2}) - \frac{l f^{(2)}(x) \kappa_{5}}{2f_{X}(x)\kappa_{2}} \nonumber \\
& & + \left\{ \frac{f^{(2)}(x)}{f_{X}(x)} - \left(\frac{f^{(1)}(x)}{f_{X}(x)}\right)^{2} \right\}\frac{l^{2}(\kappa_{2}^{2} - \kappa_{4})}{\kappa_{2}} \Biggr) m^{(2)}(x) + \frac{lf^{(1)}(x)(3l(\kappa_{2}^{2}-\kappa_{4})-\kappa_{5})}{3f_{X}(x)\kappa_{2}} m^{(3)}(x) \nonumber \\
& & + \frac{1}{2} \left[ \frac{\kappa_{4}}{6} - \frac{l \kappa_{5}}{6 \kappa_{2}} + \frac{l^{2}(3\kappa_{2}^{2}-2\kappa_{4})}{3 \kappa_{2}} - \frac{l^{4}}{2} \right] m^{(4)}(x) \Biggr ] h^{4} + O_{p} \left( r_{4} \right),
\end{eqnarray}
where $\kappa_{i} = \int t^{i} K_{X}(t) dt$ and $r_{i} = o(h^{i}) + O(h^{2}(nh)^{-1/2})$. With the choice $\pm l = \kappa_{2}^{1/2}$ in \eqref{expansion}, the SK method produces a bias with an order of magnitude $O_{p}(h^{3})$ instead of $O_{p}(h^{2})$. In addition, averaging $\widehat{m_{h}}(x|x-lh)$ and $\widehat{m_{h}}(x|x+lh)$ at $x$ can eliminate the term $h^{3}$. Hence, the convex combination (CC) of one LL estimator $\widehat{m_{h}}(x|x)$ and two SK estimators $\widehat{m_{h}}(x|x-lh)$ and $\widehat{m_{h}}(x|x+lh)$, written as
\begin{eqnarray} \label{convex_estimator}
\widehat{m_{h, \lambda}}(x) = \frac{\lambda \widehat{m_{h}}(x|x - l h) + \widehat{m_{h}}(x|x) + \lambda \widehat{m_{h}}(x|x + l h)}{1+2\lambda}, \ \mbox{where}\  \lambda > 0,
\end{eqnarray}
can simultaneously eliminate the terms $h^2$ and $h^3$ by choosing the interval parameter $l(\lambda)$, which is given by
\begin{eqnarray} \label{l(r)}
l(\lambda) = \left[ \left( 1 + \frac{1}{2\lambda} \right) \kappa_{2} \right]^{\frac{1}{2}}.
\end{eqnarray}
Thus, the asymptotic bias and variance of the CC estimator \eqref{convex_estimator} using \eqref{l(r)} are written, respectively, as
\begin{eqnarray} \label{bias_skew}
& & E_{X_{i}, Y_{i}} \left[ \widehat{m_{h, \lambda}}(x) - m(x) | X_{1}, X_{2}, ..., X_{n} \right] = B(x)h^{4} + o_{p} \left( h^{4} + \frac{1}{\sqrt{nh}} \right),
\end{eqnarray}
where
\begin{eqnarray}
B(x) = \frac{2\lambda(\kappa_{2}^{2} - \kappa_{4})\left[2f^{(2)}(x) m^{(2)}(x) + 4f^{(1)}(x)m^{(3)}(x) + f_{X}(x)m^{(4)}(x)\right] - \kappa_{2}^{2}f_{X}(x)m^{(4)}(x)}{16\lambda f_{X}(x)}, \ \ \ \ \ \ \label{term.B}
\end{eqnarray}
and
\begin{eqnarray} \label{variance_skew}
& & Var_{X_{i}, Y_{i}} \left[ \widehat{m_{h, \lambda}}(x)| X_{1}, X_{2}, ..., X_{n} \right] = \frac{1}{nh} \left[ \frac{\sigma^{2}(x)}{f_{X}(x)} \right] V(\lambda) + o_{p} \left( \frac{1}{nh} \right), \\
\mbox{where} \nonumber \\
& & V(\lambda) = \frac{1}{(2\lambda+1)^2} \left[ (2\lambda^2 + 1) \IntKer2 + (6 \lambda + 1) \int K_{X}(t-l) K_{X}(t)dt \right. \nonumber \\
& & \left. \ \ \ + \frac{(4\lambda+1)^2}{2} \int K_{X}(t-l) K_{X}(t+l)dt + \frac{\lambda(2\lambda+1)}{\kappa_{2}} \int t^{2} \left[ K_{X}^{2}(t) - K_{X}(t-l) K_{X}(t+l) \right] dt \right] \ \ \ \ \ \label{term.V} \\
\mbox{and} \nonumber \\
& & \IntKer2 = \int K_{X}^{2}(t) dt. \nonumber
\end{eqnarray}

A problem with nonparametric regression estimators is that the estimator variance \\
$Var_{X_{i}, Y_{i}}[\widehat{m_{h, \lambda}}(x)]$ is not constant over its domain because the term $[\sigma^{2}(x)/f_{X}(x)]$ appears in the leading term of \eqref{variance_skew}. To make the estimator variance approximately constant over all values of the regressor variable, Nishida and Kanazawa (2011, 2015) proposed a variance-stabilizing (VS) local variable bandwidth for the univariate Nadaraya-Watson regression estimator. The proposed VS bandwidth assumes a class of local variable bandwidths, $h_{VS}(x) = [\sigma^{2}(x)/f_{X}(x)] h_{0}$, where the global parameter $h_{0}$ is determined to minimize the asymptotic mean integrated squared error (AMISE). In the case of the CC estimator, the VS bandwidth is
\begin{eqnarray} \label{h.VS.nishida}
h_{VS}(x) &=& \frac{\sigma^{2}(x)}{f_{X}(x)} \cdot V^{\frac{1}{9}}(\bar{\lambda}_{VS}) \left[8\int_{I} \frac{\sigma^{16}(x) B^{2}(x)}{f_{X}^{7}(x)} dx \right]^{-\frac{1}{9}} \cdot n^{-\frac{1}{9}}
\end{eqnarray}
and the corresponding AMISE is
\begin{eqnarray} \label{AMISE.VS.nishida}
\lefteqn{AMISE \left( m(\cdot), \scalebox{2.6}[1.3]{$\widehat{\qquad}$} \hspace{-21mm} {m}_{h_{VS}(x), \bar{\lambda}_{VS}}(\cdot) \right)} \nonumber \\
&=& \left( 8^{\frac{1}{9}} + 8^{-\frac{8}{9}} \right) V^{\frac{8}{9}}(\bar{\lambda}_{VS}) \left[ \int_{I} \frac{\sigma^{16}(x) B^{2}(x)}{f_{X}^{7}(x)} dx \right]^{\frac{1}{9}} \cdot n^{-\frac{8}{9}}. \ \ \ 
\end{eqnarray}
The constant weight $\bar{\lambda}_{VS}$ in \eqref{h.VS.nishida} and \eqref{AMISE.VS.nishida} can be one of two types, $\bar{\lambda}_{VS, Var}$ minimizing the constant estimator variance or $\bar{\lambda}_{VS, MISE}$ minimizing mean integrated squared error (MISE). This VS bandwidth can never outperform the mean squared error (MSE)-minimizing bandwidth in terms of MISE in a univariate setting. Nishida and Kanazawa (2015) also proposed a VS local variable diagonal bandwidth matrix for a multivariate LL estimator and presented the sufficient condition under which the VS bandwidth matrix can outperform the MSE-minimizing diagonal bandwidth matrix in terms of the MISE.

In contrast, for the CC estimator $\widehat{m_{h, \lambda}}(x)$, we can propose another strategy for achieving homoscedasticity that involves controlling the weighting parameter $\lambda$ in \eqref{convex_estimator}, which is essentially dependent on $x$. Although Choi and Hall (1998) suggested that $\lambda(x)$ should be determined to minimize the term $V(\lambda(x))$ in \eqref{variance_skew} grobally, we set $\lambda(x) = \lambda_{VS}^{*}(x)$ to satisfy
\begin{eqnarray}
V(\lambda_{VS}^{*}(x)) \left[\sigma^{2}(x)/f_{X}(x) \right] = \zeta \nonumber
\end{eqnarray}
at every $x$ in a domain, where $\zeta$ is a positive constant determined to minimize the MISE or the estimator variance. The choice of $\lambda_{VS}^{*}(x)$ depends on the type of kernel and the term $\gamma^{*}(x)= \sigma^2(x) / f_{X}(x)$, which can be rewritten as the ratio of two density functions $\gamma(x) = [\sigma^2(x) / \int_{I} \sigma^2(x) dx]/[f_{X}(x)/\int_{I} f_{X}(x) dx] = [\int_{I} f_{X}(x) dx ]/[\int_{I} \sigma^2(x) dx] \gamma^{*}(x) $. Then, the constant bandwidth that minimizes the AMISE of the CC estimator employing the weight $\lambda_{VS}^{*}(x)$ and the corresponding AMISE are, respectively,
\begin{eqnarray} \label{h.VS.var}
{h}_{VS}^{*} = 8^{-\frac{1}{9}} \left[ \frac{\zeta^{}}{\int_{I} f_{X}(x) B^{2}(x; \lambda_{VS}^{*}(x)) dx} \right]^{\frac{1}{9}} \cdot n^{-\frac{1}{9}}.
\end{eqnarray}
and
\begin{eqnarray} \label{AMISE.VS.var}
{AMISE \left( m(\cdot), \scalebox{2.7}[1.3]{$\widehat{\qquad}$} \hspace{-21mm} {m}_{{h}_{VS}^{*}, \lambda_{VS}^{*}(x)}(\cdot) \right)} = \left( 8^{\frac{1}{9}} + 8^{-\frac{8}{9}} \right) \zeta^{\frac{8}{9}} \left[ \int_{I} f_{X}(x) B^{2}(x; \lambda_{VS}^{*}(x)) dx \right]^{\frac{1}{9}} \cdot n^{-\frac{8}{9}}.
\end{eqnarray}

\begin{figure}
\begin{center}
\includegraphics[width=0.5\textwidth]{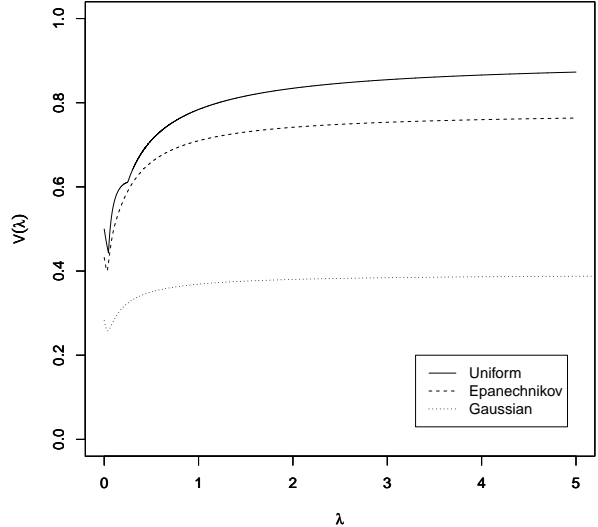}
\caption[]{The plots of $V(\lambda)$ with respect to $\lambda(x)$ when employed Unifrom, Epanechnikov and Gaussian kernels. The value $V(0)$ corresponds to $\IntKer2$.}
\label{Function.V(lambda)}
\end{center}
\end{figure}

However, variance stabilization using the weighting parameter $\lambda(x)$ poses a problem that the term $V(\lambda(x))$ in \eqref{term.V} has the upper and the lower bounds with respect to $\lambda(x)$ that are dependent on the type of kernel. Figure~\ref{Function.V(lambda)} illustrates the function $V(\lambda(x))$ with respect to $\lambda(x)$ when uniform, Epanechnikov, and Gaussian kernels are employed and represents the bounds of $V(\lambda(x))$. Following Choi and Hall (1998), the upper bound of the term $V(\lambda(x))$ is given by
\begin{eqnarray} \label{variance_skew_bound}
\frac{1}{2} \IntKer2 + 2 \int K_{X}(t - \kappa_{2}^{\frac{1}{2}})K_{X}(t + \kappa_{2}^{\frac{1}{2}}) dt + \int \frac{1}{2\kappa_{2}} t^{2} \left[ K^{2}(t) - K_{X}(t - \kappa_{2}^{\frac{1}{2}})K_{X}(t + \kappa_{2}^{\frac{1}{2}}) \right] dt. \nonumber
\end{eqnarray}
The lower bound cannot be necessarily written explicity. Since we allow the term $[\sigma^2(x)/f_{X}(x)]$ to range from $0$ to infinity in general, we cannot address every type of data; moreover, the selection of kernel function is important for variance stabilization using $\lambda(x)$.

The reminder of this paper is organized as follows. In Section~\ref{VSLL_SK}, we propose a new VS method using the weighting parameter $\lambda^{*}(x)$ in $\widehat{m_{h, \lambda}}(x)$ and explain its limitation resulting from the boundedness of $V(\lambda(x))$. We also discuss the desirable kernels for variance stabilization using the weighting method. In Section~\ref{Simulation}, we conduct simulation studies to compare two VS methods for the CC estimator, namely the VS local variable bandwidth and the VS weighting methods, because both the methods achieve homoscedasticity with regard to the leading term. This makes simulation studies indispensable for understanding their behaviors up to the remainder terms. We provide discussion and draw conclusion in Section~\ref{Discussion}.
\section{Variance-Stabilizing Weighting Method for the CC Estimator} \label{VSLL_SK}

The following proposition provides a feasibility condition for constructing a VS kernel regression estimator using the weighting method with the CC estimator \eqref{convex_estimator} and illustrates how $\lambda(x)$ can be determined. In the proposition, we define $\lambda_{max} = \arg \max_{\lambda} V(\lambda)$ and $\lambda_{min} = \arg \min_{\lambda} V(\lambda)$. We also define $x_{max} = \arg \max_{x \in I} \gamma^{*}(x)$ and $x_{min} = \arg \min_{x \in I} \gamma^{*}(x)$.
{\Proposition{\hspace{-2.0mm}{\bf{.}} \hspace{-2.0mm} The CC estimator produces a constant estimator variance when the weighting parameter $\lambda_{VS}^{*}(x)$ in \eqref{convex_estimator} is determined to satisfy $V(\lambda_{VS}^{*}(x)) \gamma^{*}(x) = \zeta$, where $\zeta$ is a positive constant, at every $x$ in the domain.\\\\
(i) Variance stabilization is feasible if and only if
\begin{eqnarray} \label{feasible.VS}
\frac{\gamma^{*}(x_{max})}{\gamma^{*}(x_{min})} \le \frac{V(\lambda_{max})}{V(\lambda_{min})}.
\end{eqnarray}
(ii) To minimize the constant estimator variance, the weight $\lambda_{VS, Var}^{*}(x)$ must be employed at every $x \in I$ satisfying
\begin{eqnarray} \label{zeta.minimizing.constant.variance}
&\ & V(\lambda_{VS, Var}^{*}(x)) \gamma^{*}(x) = \zeta_{Var}^{*} \nonumber \\
&\mbox{where}& \ \ \zeta_{Var}^{*} = V(\lambda_{min}) \gamma^{*}(x_{max}).
\end{eqnarray}
\label{Prop.feasible.VS}}
(iii) To minimize MISE, the weight $\lambda^{*}_{VS, MISE}(x)$ must be employed at every $x \in I$ satisfying
\begin{eqnarray} \label{zeta.minimizing.mise}
&\ & V(\lambda_{VS, MISE}^{*}(x)) \gamma^{*}(x) = \zeta_{MISE}^{*} \ \ \ \nonumber \\
&\mbox{where}&\ \ \ \zeta_{MISE}^{*} = {\mbox{argmin}}_{\zeta} AMISE \left(m(x), \scalebox{2.7}[1.3]{$\widehat{\qquad}$} \hspace{-21mm} m_{h_{VS}^{*}, \lambda_{VS}^{*}(x)}(x) \right) \nonumber \\
\ \ \ \ \ &\mbox{s.t.}& V(\lambda_{min}) \gamma^{*}(x_{max}) \le \zeta \le V(\lambda_{max}) \gamma^{*}(x_{min}).
\end{eqnarray}}
\hspace{-0.0mm}{\bf{Proof of (i):}}
We define the function $V = \psi(\gamma^{*}) = \zeta / \gamma^{*}$, a monotonically decreasing function with respect to $\gamma^{*}$. To make the term $V(\lambda(x)) \gamma^{*}(x)$ constant over the domain, $\psi(\gamma^{*}) = \zeta / \gamma^{*}$ must be in the range $V(\lambda_{min}) \le \psi(\gamma^{*}) \le V(\lambda_{max})$ for every $\gamma^{*}$ in $\gamma^{*}(x_{min}) \le \gamma^{*} \le \gamma^{*}(x_{max})$. This is equivalent to $\psi(\gamma^{*}(x_{min})) \le V(\lambda_{max})$ and $V(\lambda_{min}) \le \psi(\gamma^{*}(x_{max}))$. This condition yields (\ref{feasible.VS}).
\hfill $\Box$\\
\hspace{-0.0mm}{\bf{Proof of (ii):}}
From (\ref{feasible.VS}), we obtain the range $V(\lambda_{min}) \gamma^{*}(x_{max}) \le \zeta \le V(\lambda_{max}) \gamma^{*}(x_{min})$. For the constant estimator variance to be minimized, the parameter $\zeta$ must be equal to the lower bound of the range.
\hfill $\Box$\\
\hspace{-0.0mm}{\bf{Proof of (iii):}}
Considering that the parameter $\lambda(x)$ is included in the denominator of the bias term $B(x)$ in (\ref{bias_skew}), we can determine the local parameter $\lambda_{VS}^{*}(x)$ and the global parameter $\zeta$ that minimize the AMISE in \eqref{AMISE.VS.var} under the constraints $V(\lambda_{VS}^{*}(x)) \gamma^{*}(x) = \zeta$ and $V(\lambda_{min}) \gamma^{*}(x_{max}) \le \zeta \le V(\lambda_{max}) \gamma^{*}(x_{min})$.
\hfill $\Box$\\
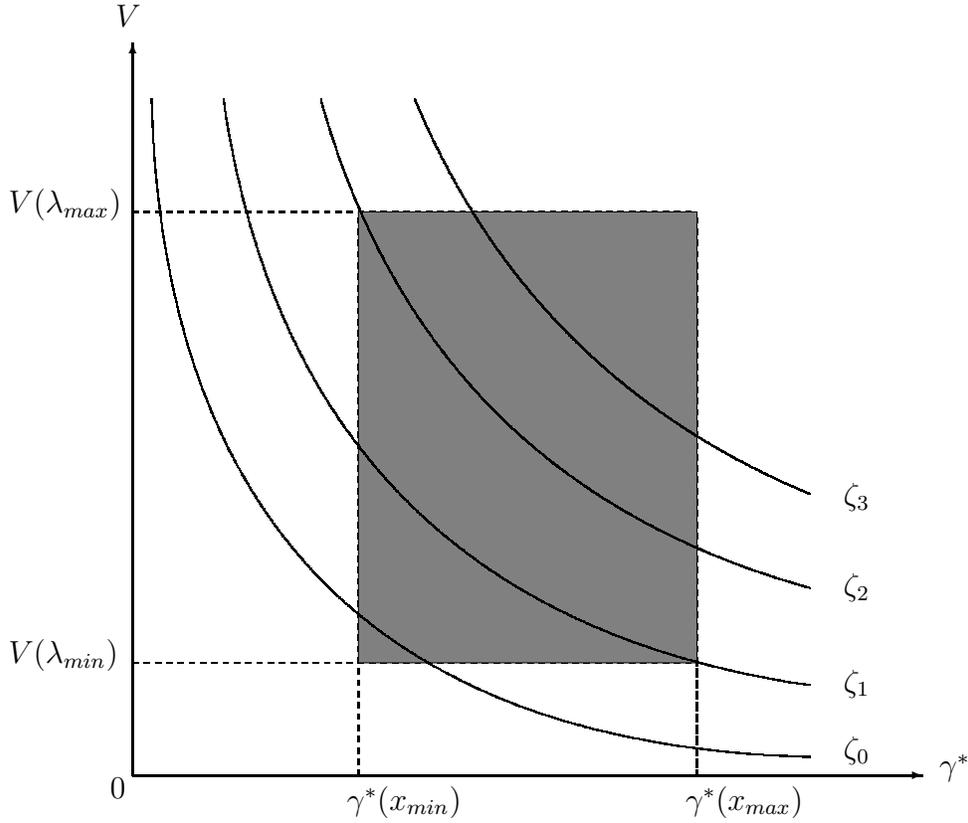
\begin{figure}
\setlength\unitlength{1.5truecm}
\begin{center}
\begin{picture}(10,10)(0,0)
\put(1,1){
\put(0, 0){\vector(1, 0){7}}
\put(0, 0){\vector(0, 1){6.5}}
\put(0, 0){\dashbox{0.05}(5, 5){}}
\put(0, 0){\dashbox{0.05}(5, 1){}}
\put(0, 0){\dashbox{0.05}(2, 5){}}
\shade \path(2, 1)(5, 1)(5, 5)(2, 5)(2, 1)
\put(-0.2, -0.2){0}
\put(7.15, 0){$\gamma^{*}$}
\put(-0.15, 6.65){$V$}
\put(1.9, -0.3){$\gamma^{*}(x_{min})$}
\put(4.9, -0.3){$\gamma^{*}(x_{max})$}
\put(-1.1, 1){$V(\lambda_{min})$}
\put(-1.1, 5){$V(\lambda_{max})$}
\put(0, 0){\qbezier[1000](0.1666, 6)(0.3243, 0.3243)(6, 0.1666)}
\put(0, 0){\qbezier[1000](0.80333, 6)(1.46341, 1.46341)(6, 0.80333)}
\put(0, 0){\qbezier[1000](1.6667, 6)(2.6087, 2.6087)(6, 1.6667)}
\put(0, 0){\qbezier[1000](2.5, 6)(3.52941, 3.52941)(6, 2.5)}
\put(6.3, 0.15){$\zeta_{0}$}
\put(6.3, 0.75){$\zeta_{1}$}
\put(6.3, 1.6){$\zeta_{2}$}
\put(6.3, 2.40){$\zeta_{3}$}
}
\end{picture}
\end{center}
\caption[]{Diagram to illustrate the proof of proposition~1. We can choose any combination $(\gamma^{*}(x), V(\lambda))$ in the shaded area of the diagram. We notice that $\zeta_{0} < \zeta_{1} < \zeta_{2} < \zeta_{3}$, and the values of the parameters $\zeta_{1}$ and $\zeta_{2}$ are obtained when $(\gamma^{*}(x), V(\lambda))$ is $(\gamma^{*}(x_{max}), V(\lambda_{min}))$ and $(\gamma^{*}(x_{min}), V(\lambda_{max}))$ respectively. Variance stabilization using the weighting parameter $\lambda(x)$ is feasible, if $\zeta_{1} < \zeta < \zeta_{2}$. This condition is equivalent to \eqref{feasible.VS}. The estimator variance is minimized when $\zeta_{Var}^{*} = \zeta_{1}$. The MISE will be minimized under the constraint $\zeta_{1} < \zeta < \zeta_{2}$.} \label{Proof.Diagram}
\end{figure}

Hereafter, we denote $\zeta_{MISE}^{*}$ and $\zeta_{Var}^{*}$ to be the values of the parameter $\zeta$'s that minimize the MISE and the estimator variance, respectively. To help understand the proof, please refer to Figure~\ref{Proof.Diagram}. We also present the following illustrative example to see in which cases variance stabilization using the weighting method is feasible.\\
\\{\bf{Example 1.}} Suppose that we employ the Gaussian kernel. Then, we obtain $V(\lambda_{max}) / V(\lambda_{min}) = 1.5223$ and $V(\lambda_{max}) - V(\lambda_{min}) = 0.1351$. Consider the following two cases.\\
(i) $f_{X}(x) = (1/0.3829)(1/\sqrt{2\pi}) \exp \left(-0.5(x-0.5)^2 \right)$ and $\sigma^2(x) = 2.5 + |x - 0.5|$ defined in $I=[0, 1]$. In this case, $\gamma^{*}(x_{max}) - \gamma^{*}(x_{min}) = 3.2629 - 2.3996 = 0.8669$. Then, we obtain $\gamma^{*}(x_{max})/\gamma^{*}(x_{min}) = 1.3597 < 1.5236 = V(\lambda_{max}) / V(\lambda_{min})$, and we notice that variance stabilization is feasible. The range of $\zeta$ is $[0.8418, 0.9432]$, and the minimized constant estimator variance is $V(\lambda_{min}) \gamma^{*}(x_{max}) (1/nh) = 0.8418 (1/nh)$. \\
(ii) $f_{X}(x) = (1/0.3829)(1/\sqrt{2\pi}) \exp \left(-0.5(x-0.5)^2 \right) $ and $\sigma^2(x) = 0.05 + |x - 0.5|$ defined in $I=[0, 1]$. In this case, $\gamma^{*}(x_{max}) - \gamma^{*}(x_{min}) = 0.0815 - 0.0479 = 0.0336$. Then, we obtain $\gamma^{*}(x_{max})/\gamma^{*}(x_{min}) = 1.7014 > 1.5236 = V(\lambda_{max}) / V(\lambda_{min})$. Variance stabilization is infeasible.\\
{\Remark{{\hspace{-2mm}\bf{.}} {\rm{To decrease the constant estimator variance, it is desirable that we choose kernels whose $V(\lambda_{min})$ values are small and/or that we deal with data containing small values of $\gamma^{*}(x_{max})$.}}} \label{Proposition_Remark_1}}
{\Remark{{\hspace{-2mm}\bf{.}} {\rm{It is possible that there exist multiple weighting parameters $\lambda_{VS}^{*}(x)$ satisfying $V(\lambda_{VS}^{*}(x)) \gamma^{*}(x) = \zeta^{*}$ for a $x$. In such cases, we employ the largest parameter to assure continuity of $\lambda_{VS}^{*}(x)$. }}} \label{Two_intersections}}
\\\\
{\bf{The algorithm}}

We summarize the algorithm for variance stabilization using the weighting parameter $\lambda_{VS}^{*}(x)$ in $\widehat{m_{h, \lambda}}(x)$.
\begin{enumerate}
\item Choose a kernel, and calculate $V(\lambda_{max})/V(\lambda_{min})$.
\item Estimate $\widehat{\sigma^2}(x)/\widehat{f_{X}}(x)$ and calculate $\widehat{\gamma^{*}}(x_{max})/\widehat{\gamma^{*}}(x_{min})$ in $x \in I$.
\item If $\widehat{\gamma^{*}}(x_{max})/\widehat{\gamma^{*}}(x_{min}) > V(\lambda_{max})/V(\lambda_{min})$, then variance stabilization is not feasible. Otherwise, go to the next step.
\item Calculate $\lambda_{VS}^{*}(x)$ satisfying $V(\lambda_{VS}^{*}(x)) \left[\widehat{\sigma^{2}}(x)/\widehat{f_{X}}(x) \right] = \zeta^{*} > 0$ at every $x$ in $I$.
\begin{enumerate}[i.]
\item To minimize the constant estimator variance, $\zeta^{*}$ is set to be $\zeta_{Var}^{*}$.
\item To minimize MISE, $\zeta^{*}$ is set to be $\zeta^{*}_{MISE}$.
\end{enumerate}
\item Estimate $\widehat{B}(x)$ in \eqref{term.B} with $\lambda_{VS}^{*}(x)$. Then, obtain the estimator $\widehat{h_{VS}^{*}}$ from \eqref{h.VS.var}.
\item Calculate $l(\lambda_{VS}^{*}(x))$ at every $x$ in $I$. 
\item Construct $\widehat{m_{h, \lambda}}(x)$ such that $\lambda(x) = \lambda_{VS}^{*}(x)$, $l(\lambda(x)) = l(\lambda_{VS}^{*}(x))$ and $h=\widehat{h_{VS}^{*}}$.
\end{enumerate}
{\bf{Choice of Kernel}}

\begin{figure}
\begin{center}
\includegraphics[width=0.45\textwidth]{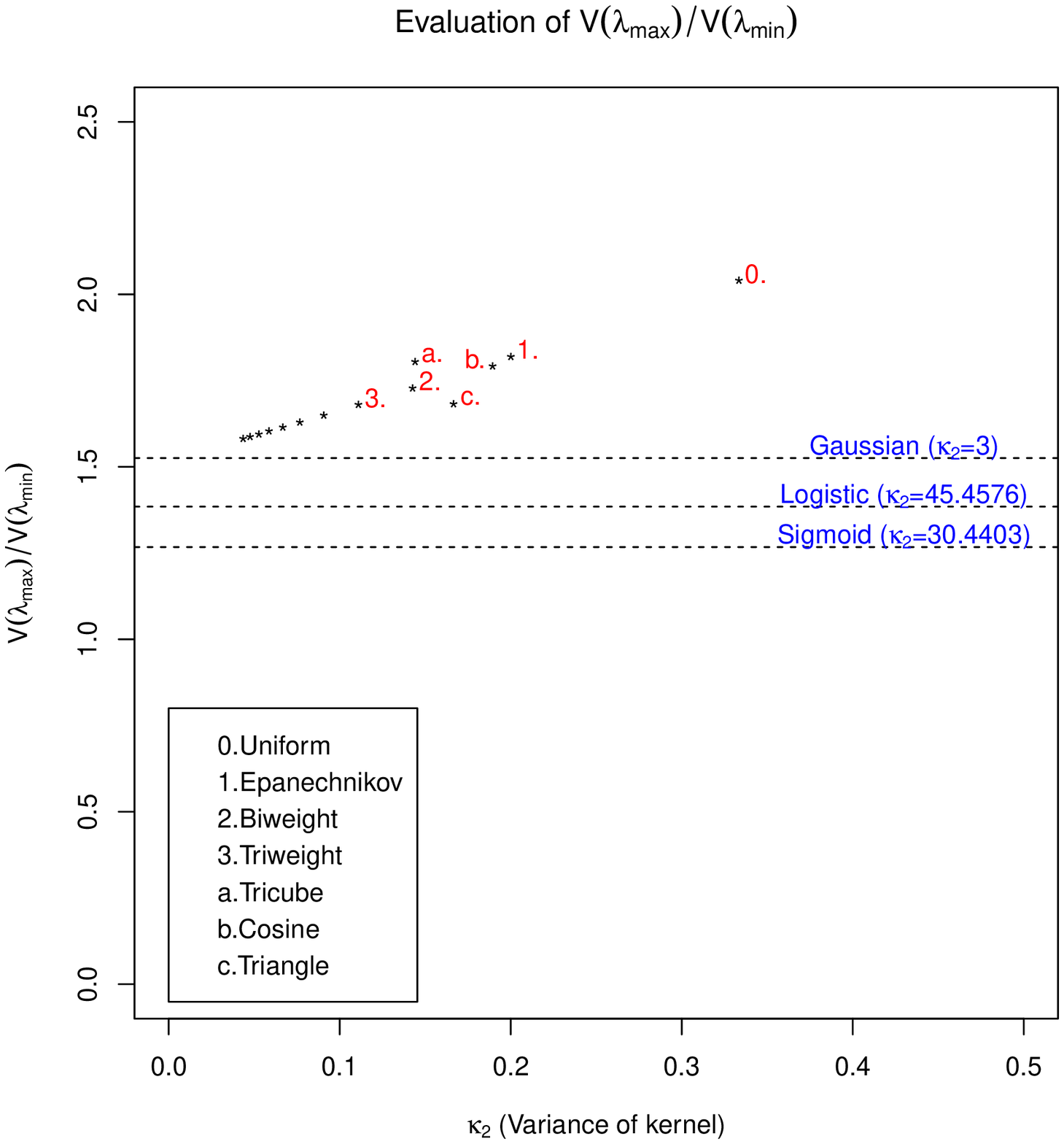}
\includegraphics[width=0.45\textwidth]{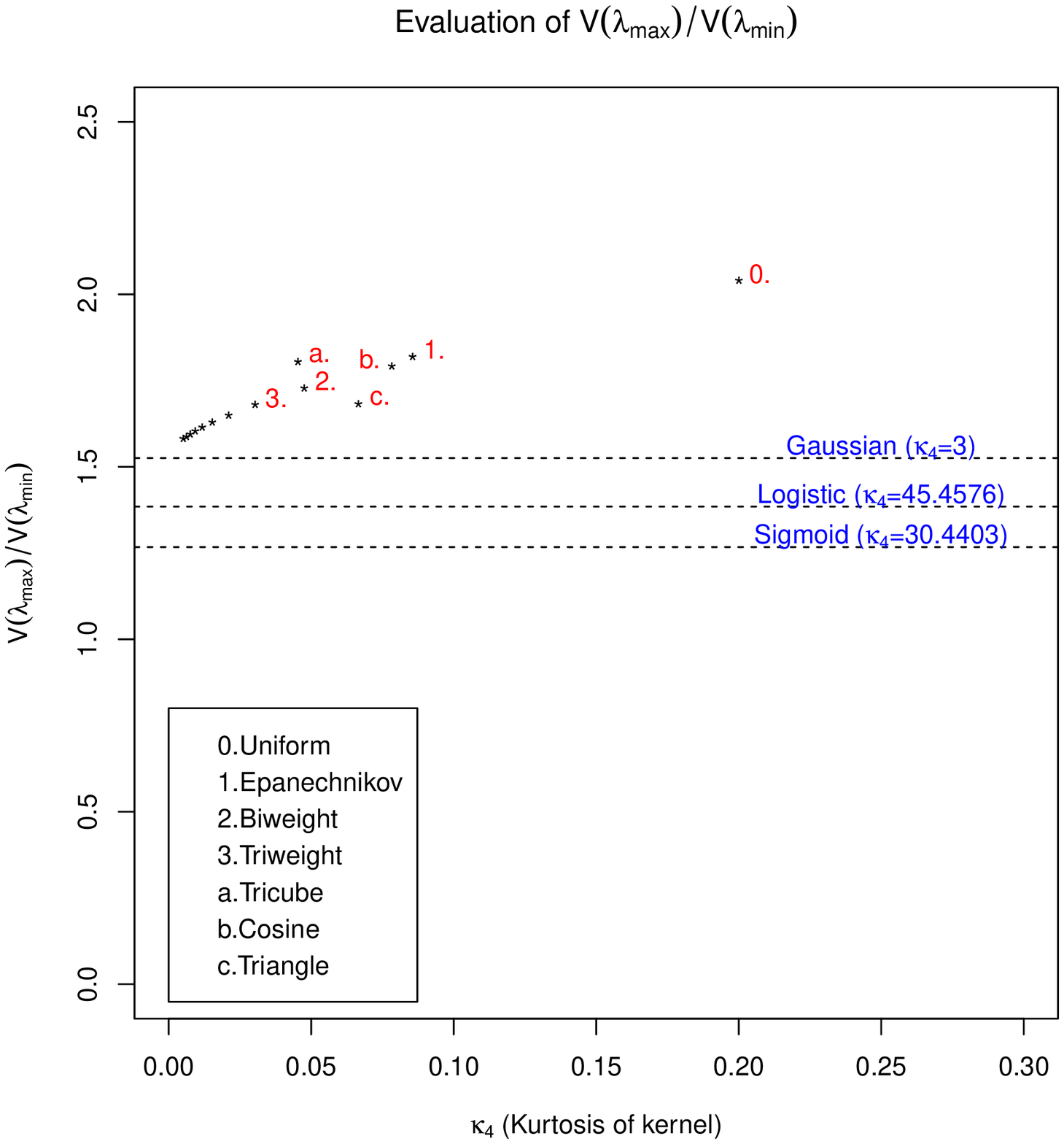}
\caption[]{Plots of the relation between the value of $V(\lambda_{max})/V(\lambda_{min})$ and the variance of the kernel $\kappa_{2}$ for each kernel (left panel). Plots of the relation between $V(\lambda_{max})/V(\lambda_{min})$ and the kurtosis of kernel $\kappa_{4}$ for each kernel (right panel).}
\label{Evaluation.V.k2.k4}
\end{center}
\end{figure}

\begin{figure}
\begin{center}
\includegraphics[width=0.45\textwidth]{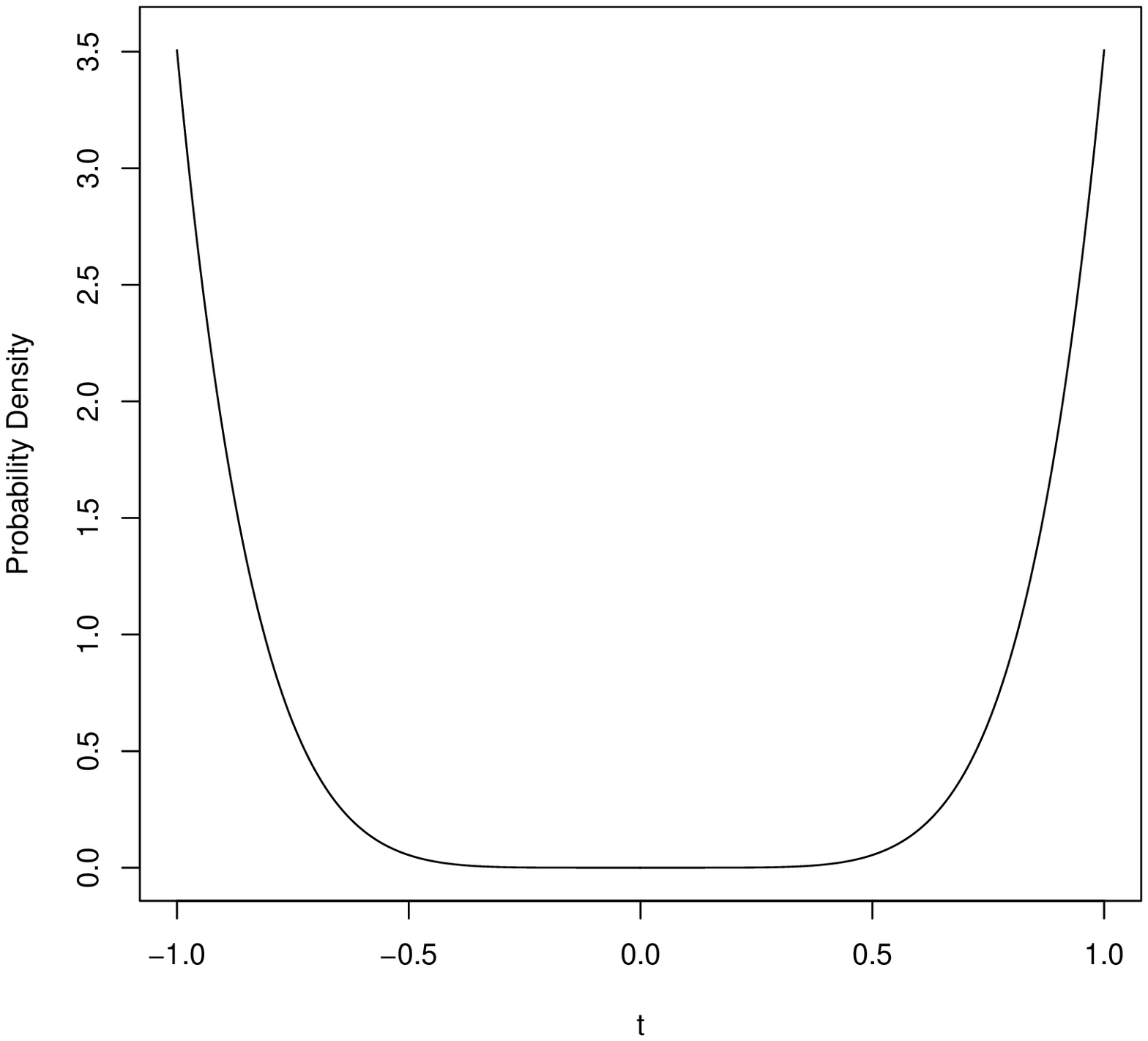}
\includegraphics[width=0.45\textwidth]{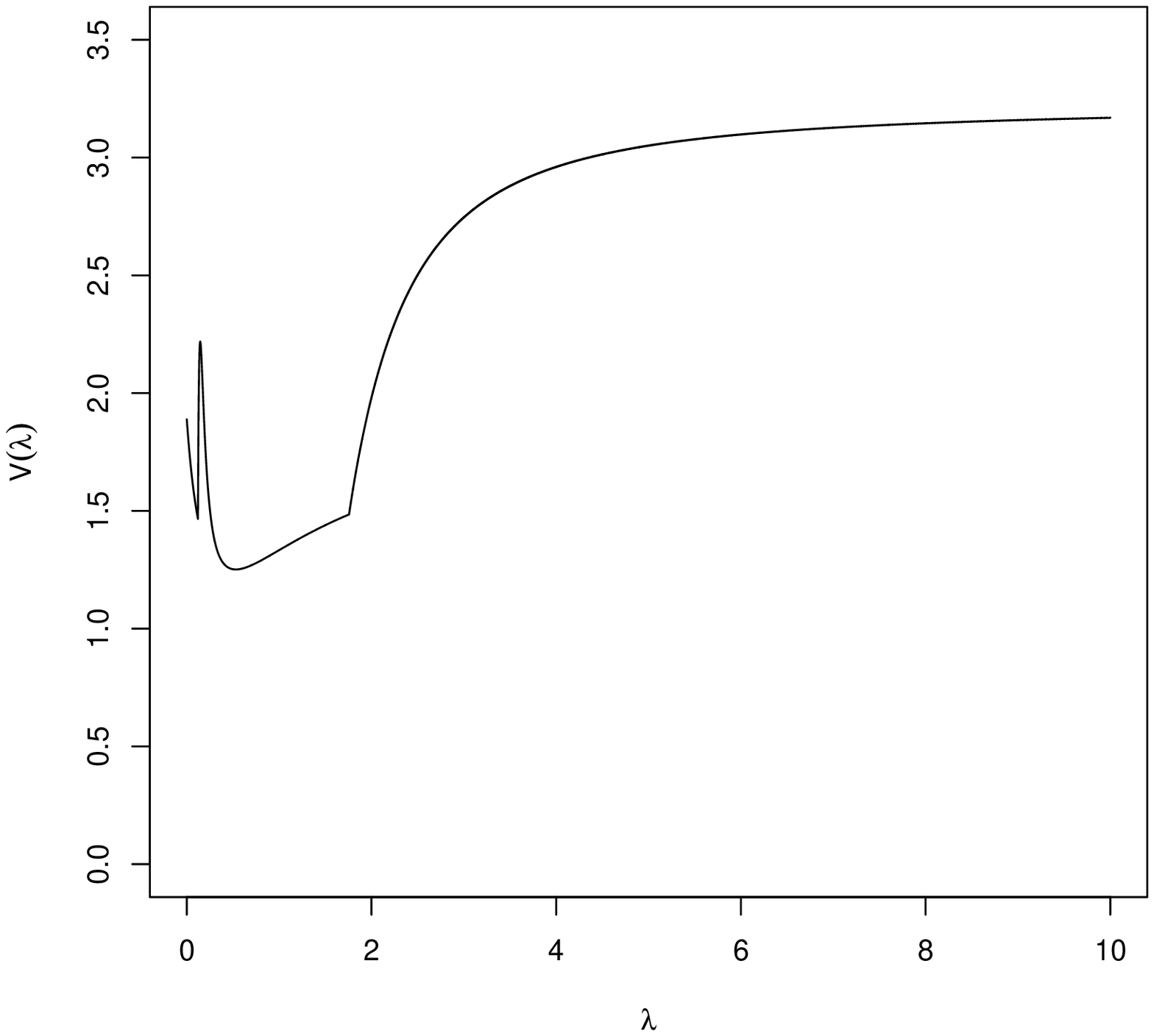}
\caption[]{Left panel: Graphic of the kernel function that maximizes $V(\lambda_{max})/V(\lambda_{min})$ for the class in \eqref{Kernel.VS} based on $K_{X}(t) = 3.5065 |t|^{6.0131} I_{[-1,1]}(t)$. Right panel: Plot of the function $V(\lambda)$ for this case: $\kappa_{2} = 0.7780$, $\kappa_{4} = 0.6367$, $V(\lambda_{min}) = 1.2533$, $V(\lambda_{max}) = 3.2484$, $V(\lambda_{max})/V(\lambda_{min}) = 2.5918$, and $\lambda_{min} = 0.5351$.}
\label{VS.Kernel}
\end{center}
\end{figure}

From \eqref{feasible.VS}, we notice that the choice of kernel is important for variance stabilization when using the weighting method because kernels that yield a larger value of $V(\lambda_{max})/V(\lambda_{min})$ can handle many types of data. To follow our discussion of this point, see Table \ref{evaluation.V}, in which we evaluate the term $V(\lambda_{max})/V(\lambda_{min})$ for different types of kernels by employing Gaussian, cosine, and triangle kernels, as well as the kernel in Wand and Jones (1995, p.31), i.e.,
\begin{eqnarray} \label{beta_kernel}
K_{X}(t) = \frac{(1-t^{2})^{\theta}}{2^{2\theta+1} \mbox{Beta}(\theta+1, \theta+1)}I_{[-1, 1]}(t), \ \ \theta \ \mbox{is a positive integer},
\end{eqnarray}
for $\theta = 0, 1, 2, ..., 10$, which yields uniform, Epanechnikov, biweight, and triweight kernels respectively, when $\theta = 0, 1, 2$, and $3$. The function $I_{[a, b]}( \cdot )$ is an indicator function defined in $[a, b]$. Fugure \ref{Evaluation.V.k2.k4} summarizes the results in the table. Those results show that the value $V(\lambda_{max})/V(\lambda_{min})$ increases as the variance or kurtosis of the kernel increases, with the exception of kernels with an unbounded domain such as Gaussian, logistic, and sigmoid. Specifically, uniform kernel yields the largest value of $V(\lambda_{max})/V(\lambda_{min})$ and is the best for the purpose of variance stabilization among the kernels presented. We also notice that kernels with unbounded domains are inadvisable.
\begin{table}
\begin{center}
{\scriptsize{
\begin{tabular}{lllllllll}
\hline
\hline
$\theta$ (Kernel type) & $\kappa_{2}$ & $\kappa_{4}$ & $V(\lambda_{min})$ & $V(\lambda_{max})$ & $\mbox{argmin}_{\lambda}V(\lambda)$ & Range $V(\lambda)$ & $V(\lambda_{max})/V(\lambda_{min})$ & \\
\hline
Tricube & 0.1440 \ \ & 0.0455 \ \ & 0.6443 \ \ & 1.1622 \ \ & 0.0359 & 0.5179 & 1.8038 \ \ \\
Cosine & 0.1894 \ \ & 0.0787 \ \ & 0.5609 \ \ & 1.0054 \ \ & 0.0350 & 0.4444 & 1.7922 \ \ \\ 
Triangle & 0.1666 \ \ & 0.0666 \ \ & 0.6083 \ \ & 1.0227 \ \ & 0.0343 & 0.4143 & 1.6810 \ \ \\
Gaussian * & 1 \ \ & 3 \ \ & 0.2580 \ \ & 0.3931 \ \ & 0.0376 & 0.1351 & 1.5223 \ \ \\
Logistic * & 3.2899 \ \ & 45.4576 \ \ & 0.1531 \ \ & 0.2120 \ \ & 0.0385 & 0.0589 & 1.3845 \ \ \\
Sigmoid * & 2.4674 \ \ & 30.4403 \ \ & 0.1856 \ \ & 0.2352 \ \ & 0.0456 & 0.0496 & 1.2672 \ \ \\
\hline
$\theta=0$ (Uniform) & 0.3333 \ \ & 0.2 \ \ & 0.4432 \ \ & 0.9037 \ \ & 0.0454 & 0.4605 & 2.0392 \ \ \\
$\theta=1$ (Epanechnikov) & 0.2 \ \ & 0.0857 \ \ & 0.5449 \ \ & 0.9914 \ \ & 0.0352 & 0.4465 & 1.8195 \ \ \\
$\theta=2$ (Biweight) & 0.1429 \ \ & 0.0476 \ \ & 0.6512 \ \ & 1.1255 \ \ & 0.0352 & 0.4743 & 1.7283 \ \ \\
$\theta=3$ (Triweight) & 0.1111 \ \ & 0.0303 \ \ & 0.7447 \ \ & 1.2509 \ \ & 0.0355 & 0.5061 & 1.6796 \ \ \\
$\theta=4$ & 0.0909 \ \ & 0.0210 \ \ & 0.8284 \ \ & 1.3665 \ \ & 0.0358 & 0.5381 & 1.6495 \ \ \\
$\theta=5$ & 0.0769 \ \ & 0.0153 \ \ & 0.9047 \ \ & 1.4738 \ \ & 0.0360 & 0.5691 & 1.6290 \ \ \\
$\theta=6$ & 0.0666 \ \ & 0.0117 \ \ & 0.9752 \ \ & 1.5743 \ \ & 0.0361 & 0.5990 & 1.6142 \ \ \\
$\theta=7$ & 0.0588 \ \ & 0.0093 \ \ & 1.0410 \ \ & 1.6689 \ \ & 0.0363 & 0.6278 & 1.6030 \ \ \\
$\theta=8$ & 0.0526 \ \ & 0.0075 \ \ & 1.1030 \ \ & 1.7585 \ \ & 0.0364 & 0.6555 & 1.5943 \ \ \\
$\theta=9$ & 0.0476 \ \ & 0.0062 \ \ & 1.1617 \ \ & 1.8440 \ \ & 0.0365 & 0.6822 & 1.5873 \ \ \\
$\theta=10$ & 0.0435 \ \ & 0.0052 \ \ & 1.2175 \ \ & 1.9257 \ \ & 0.0366 & 0.7082 & 1.5816 \ \ \\
\hline
\hline
\end{tabular}}}
\caption[]
{Evaluation of the term $V(\lambda)$ for different types of kernels. Kernels with infinite domains are marked with $*$. Gaussian: $K_{X}(t) = \frac{1}{\sqrt{2\pi}} \exp{\left(-\frac{t^2}{2}\right)}$, Logistic: $K_{X}(t) = \frac{1}{2+\exp{(t)}+\exp{(-t)}}$, Sigmoid: $K_{X}(t) = \frac{2}{\pi} \left[ \frac{1}{\exp{(t)}+\exp{(-t)}} \right]$.
} \label{evaluation.V}
\end{center}
\end{table}
\clearpage
{\Remark{{\hspace{-2mm}\bf{.}} {\rm{Let us consider the following kernel function}}} \label{VS_Kernel}}
\begin{eqnarray} \label{Kernel.VS}
K_{X}(t) = \left[ a_{0} + \frac{1}{2}(1-2 a_{0})(a_{1}+1)t^{a_{1}} \right] I_{[-1, 1]}(t),\ \ \ a_{1} > 0, \ \ \frac{1+a_{1}}{2 a_{1}} > a_{0} > 0,
\end{eqnarray}
which yields uniform, triangle, and Epanechnikov kernels, respectively, when $(a_{0}, a_{1}) = (1/2, \mbox{any value})$, $(1, 1)$ and $(3/4, 2)$. When $0 < a_{0} < 1/2$, the kernel function does not satisfy the essential property for kernels that the point at which we want to estimate the probability density must decrease (resp., increase) weight if the point is located farther from (resp., closer to) the data point; nevertheless, it is a probability density function satisfying the standard conditions required for kernel smoothing, symmetry, nonnegativity and adequate smoothness. Considering this example and choosing $(a_{0}, a_{1}) = (0, 6.0131)$, we obtain the kernel $K_{X}(t) = 3.5065 |t|^{6.0131}I_{[-1,1]}(t)$ illustrated in the left panel of Figure.\ref{VS.Kernel}, which yields the maximim $V(\lambda_{max})/V(\lambda_{min}) = 2.5854$ among the class of kernel functions \eqref{Kernel.VS}. This value is $26.78\%$ larger than that obtained using the uniform kernel. We conjecture that the mechanism for distributing the higher weight to a point farther away is necessary for the kernel to increase the value of $V(\lambda_{max})/V(\lambda_{min})$.
\section{Simulation} \label{Simulation}

We are especially interested in knowing to what degree the proposed methods---the VS local variable bandwidth and VS weighting methods---can stabilize the estimator variance up to the remainder terms. We are also interested in evaluating the cost of variance stabilization in terms of the MISE. For this purpose, we conduct simulations and comparisons between two proposed methods. For the sake of comparison, we employ two more methods that produce heteroscedastic regression estimators. One is the MISE-minimizing fixed bandwidth method:
\begin{eqnarray} \label{h.MISE.opt}
{h}_{fixed} = \left[ \frac{\int_{I} \sigma^{2}(x) V(\bar{\lambda}_{fixed}) dx}{\int_{I} 8 f_{X}(x) B^{2}(x) dx} \right]^{\frac{1}{9}} \cdot n^{-\frac{1}{9}},
\end{eqnarray}
with the corresponding AMISE being
\begin{eqnarray} \label{AMISE.MISE.opt}
\lefteqn{AMISE \left( m(\cdot), \scalebox{2.7}[1.3]{$\widehat{\qquad}$} \hspace{-21mm} {m}_{{h}_{fixed}, \bar{\lambda}_{fixed}}(\cdot) \right)} \nonumber \\
&=& \left( 8^{\frac{1}{9}} + 8^{-\frac{8}{9}} \right) \left[ \int_{I} \sigma^{2}(x) V(\bar{\lambda}_{fixed}) dx \right]^{\frac{8}{9}} \left[ \int_{I} f_{X}(x) B^{2}(x) dx \right]^{\frac{1}{9}} \cdot n^{-\frac{8}{9}}. \nonumber
\end{eqnarray}
The other is the MSE-minimizing local variable bandwidth method:
\begin{eqnarray} \label{h.MSE.opt}
h_{MSE}(x) = 8^{-\frac{1}{9}} \left[ \frac{\sigma^{2}(x) V(\lambda_{MSE}(x))}{f_{X}(x) B^{2}(x)} \right]^{\frac{1}{9}} \cdot n^{-\frac{1}{9}},
\end{eqnarray}
with the corresponding asymptotic mean squared error (AMSE) and AMISE being, respectively
\begin{eqnarray} \label{MSE.opt}
AMSE \left( m(\cdot), \scalebox{2.7}[1.3]{$\widehat{\qquad}$} \hspace{-21mm} {m}_{h_{MSE}, \lambda_{MSE}}(\cdot) \right) = \left( 8^{\frac{1}{9}} + 8^{-\frac{8}{9}} \right) \left[ \sigma^{\frac{16}{9}}(x) f_{X}^{-\frac{8}{9}}(x) V^{\frac{8}{9}}(\lambda_{MSE}(x)) B^{\frac{2}{9}}(x) \right] \cdot n^{-\frac{8}{9}} \nonumber
\end{eqnarray}
and
\begin{eqnarray} \label{AMISE.MSE.opt}
\lefteqn{AMISE \left( m(\cdot), \scalebox{2.7}[1.3]{$\widehat{\qquad}$} \hspace{-21mm}{m}_{h_{MSE}, \lambda_{MSE}}(\cdot) \right) = \int_{I} AMSE \left( m(\cdot), \scalebox{2.7}[1.3]{$\widehat{\qquad}$} \hspace{-21mm} {m}_{h_{MSE}, \lambda_{MSE}}(\cdot) \right) f_{X}(x) dx} \nonumber \\
&=& \left( 8^{\frac{1}{9}} + 8^{-\frac{8}{9}} \right) \left[ \int_{I} \sigma^{\frac{16}{9}}(x) f_{X}^{\frac{1}{9}}(x) V^{\frac{8}{9}}(\lambda_{MSE}(x)) B^{\frac{2}{9}}(x) dx \right] \cdot n^{-\frac{8}{9}}. \nonumber
\end{eqnarray}
With this as background, we now compare the following eight types of methods. The estimators $(a)$, $(b)$, $(e)$, and $(f)$ are homoscedastic, while $(c)$, $(d)$, $(g)$, and $(h)$ are heteroscedastic. We also form two groups, group [I] comprising $(a), (b), (c)$, and $(d)$ minimizing the estimator variance and group [II] comprising $(e), (f), (g)$, and $(d)$ minimizing the MISE; then, we make comparisons in each group.
\begin{enumerate}[(a)]
\item {\bf{VS weighting method minimizing the estimator variance:}} The CC estimator employs the weight $\lambda_{VS, Var}^{*}(x)$ in \eqref{zeta.minimizing.constant.variance}. We denote the corresponding constant bandwidth in \eqref{h.VS.var} as $h_{VS, Var}^{*}$.
\item {\bf{VS local variable bandwidth minimizing the estimator variance:}} The CC estimator employs the VS local variable bandwidth $h_{VS}(x)$ in \eqref{h.VS.nishida} with the constant weight $\bar{\lambda}_{VS} = \bar{\lambda}_{VS, Var} = \lambda_{min}$ minimizing the estimator variance. We denote the corresponding local variable bandwidth as $h_{VS, Var}(x)$.
\item {\bf{MISE-minimizing fixed bandwidth with the weighting parameter minimizing the estimator variance:}} The CC estimator employs the MISE-minimizng fixed bandwidth $h_{fixed}$ in \eqref{h.MISE.opt} with the constant weight $\bar{\lambda}_{fixed} = \bar{\lambda}_{fixed, Var} = \lambda_{min}$ minimizing the estimator variance. We denote the corresponding constant bandwidth as $h_{fixed, Var}$.
\item {\bf{MSE-minimizing bandwidth with the weighting parameter minimizing estimator variance:}} The CC estimator employs the MSE-minimizing local variable bandwidth \eqref{h.MSE.opt} and the weighting parameter $\lambda_{MSE}(x) = \lambda_{MSE, Var}(x) = \lambda_{min}$  minimizing the estimator variance. We denote the corresponding local variable bandwidth as $h_{MSE, MISE}(x)$.
\item {\bf{VS weighting method minimizing the MISE:}} The CC estimator employs the weight $\lambda_{VS, MISE}^{*}(x)$ in \eqref{zeta.minimizing.mise}. We denote the corresponding constant bandwidth in \eqref{h.VS.var} as $h_{VS, MISE}^{*}$.
\item {\bf{VS local variable bandwidth minimizing the MISE:}} The CC estimator employs the VS local variable bandwidth $h_{VS}(x)$ in \eqref{h.VS.nishida} with the constant weight $\bar{\lambda}_{VS} = \bar{\lambda}_{VS, MISE}$ minimizing the MISE. We denote the corresponding local variable bandwidth as $h_{VS, MISE}(x)$.
\item {\bf{MISE-minimizing fixed bandwidth with the weighting parameter minimizing the MISE:}} The CC estimator employs the MISE-minimizng fixed bandwidth $h_{fixed}$ in \eqref{h.MISE.opt} with the constant weight $\bar{\lambda}_{fixed} = \bar{\lambda}_{fixed, MISE}$ minimizing the MISE. We denote the corresponding constant bandwidth as $h_{fixed, MISE}$.
\item {\bf{MSE-minimizing bandwidth with weighting parameter minimizing MISE:}} The CC estimator employs the MSE-minimizing local variable bandwidth \eqref{h.MSE.opt} and the local variable weighting parameter $\lambda_{MSE}(x) = \lambda_{MSE, MISE}(x)$ minimizing MSE. We denote the corresponding local variable bandwidth as $h_{MSE, MISE}(x)$.
\end{enumerate}

The simulation setting is as follows. For the true regression function, we employ the function given in Choi and Hall (1998), $m(x) = m_{k}(x) = \frac{2}{5} \left[ 3 \sin(2k\pi x) + 2 \sin(3 \pi x) \right]$, $k=1,2,3$, defined in $I=[0, 1]$. If $k$ is greater, the frequency of the curve $y=m(x)$ becomes high. We use the settings of $f_{X}(x)$ and $\sigma^{2}(x)$ given in Example~1-(i). For the kernel, we employ a Gaussian kernel. We repeat the following process $M=100$ times at points from $0.000$ to $1.000$ with an increment of $0.001$ for $n=100$, $500$, and $1,000$. 
\\\\{\bf{Process}}
\begin{enumerate}[Step 1]
\setlength{\itemsep}{-2mm}
\item Generate $X_{i}$ with a sample size $n$ distributed as $f_{X_{i}}(x_{i})$.
\item Generate $U_{i}|\{X_{i} = x_{i} \}$ with a sample size $n$ distributed as $N(0, \sigma^2(x_{i}))$.
\item Obtain $(X_{i}, Y_{i})$ with a sample size $n$, where $Y_{i} = m(x_{i}) + U_{i}|\{X_{i} = x_{i} \}$.
\item Construct each estimator $(a)$--$(h)$ at every point $x = 0.00 + \epsilon \cdot j$, $\epsilon = 0.001$, $j = 1,..., 1000$, using the sample $(X_{i}, Y_{i})$, $i=1,...,n$, obtained in Step $1$-$3$ above.
\item Repeat Step 1 through 4 $M=100$ times.
\item Let $\widehat{m_{h, \lambda}}^{(\bf{T})}(x)$ be the CC estimator calculated $(\bf{T})$-th generated sample of size $n$. At every point $x = 0.0 + \epsilon \cdot j$, $j = 1,..., 1000$, compute the sample variances of $\widehat{m_{h, \lambda}}^{(\bf{T})}(x)$, $\mathbf{T} = 1, ..., M$, that are respectively calculated in Step $1$-$5$. Perform this computation for each estimator $(a)$-$(h)$.
\item For each estimator $(a)$-$(h)$, obtain the distribution of the sample variances of $\widehat{m_{h, \lambda}}^{(\bf{T})}(x)$, $\mathbf{T} = 1, ..., M$, calculated at $1000$ points in Step 6. Calculate the standard deviations (SDs) of the distribution for each.
\item For each estimator, compute the estimator of the MISE given by 
\begin{eqnarray} \label{estimator.MISE}
\sum_{(\bf{T})=1}^{M} \int_{I} f_{X}(x) \left[ m(x) - \widehat{m_{h, \lambda}}^{(\bf{T})}(x) \right]^{2} dx.
\end{eqnarray}
\end{enumerate}

The simulation programs are written in the programing language C and compiled using Open Watcom Version 1.5. We assume an error term $U_{i}|X_{i}$ is a normally distributed random variable with a mean of $0$ and variance $\sigma^2(X_{i})$. To generate random numbers $X_{i}$ and $U_{i}|X_{i}$, $i=1,...,n$, the algorithm in William et al. (1992, p.280) for Box-Muller method is used.

\begin{figure}
\begin{center}
\includegraphics[width=0.48\textwidth]{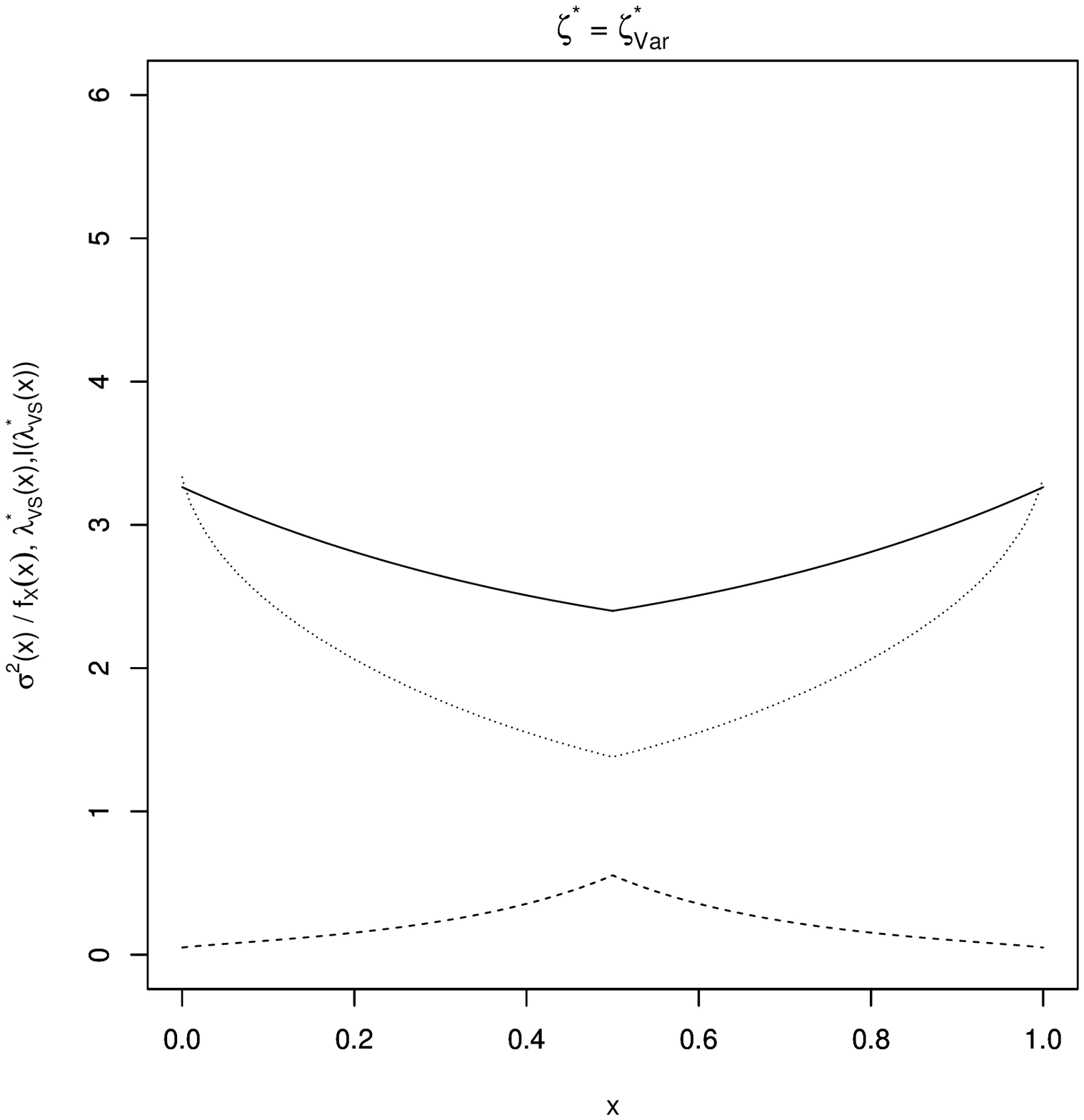}
\includegraphics[width=0.48\textwidth]{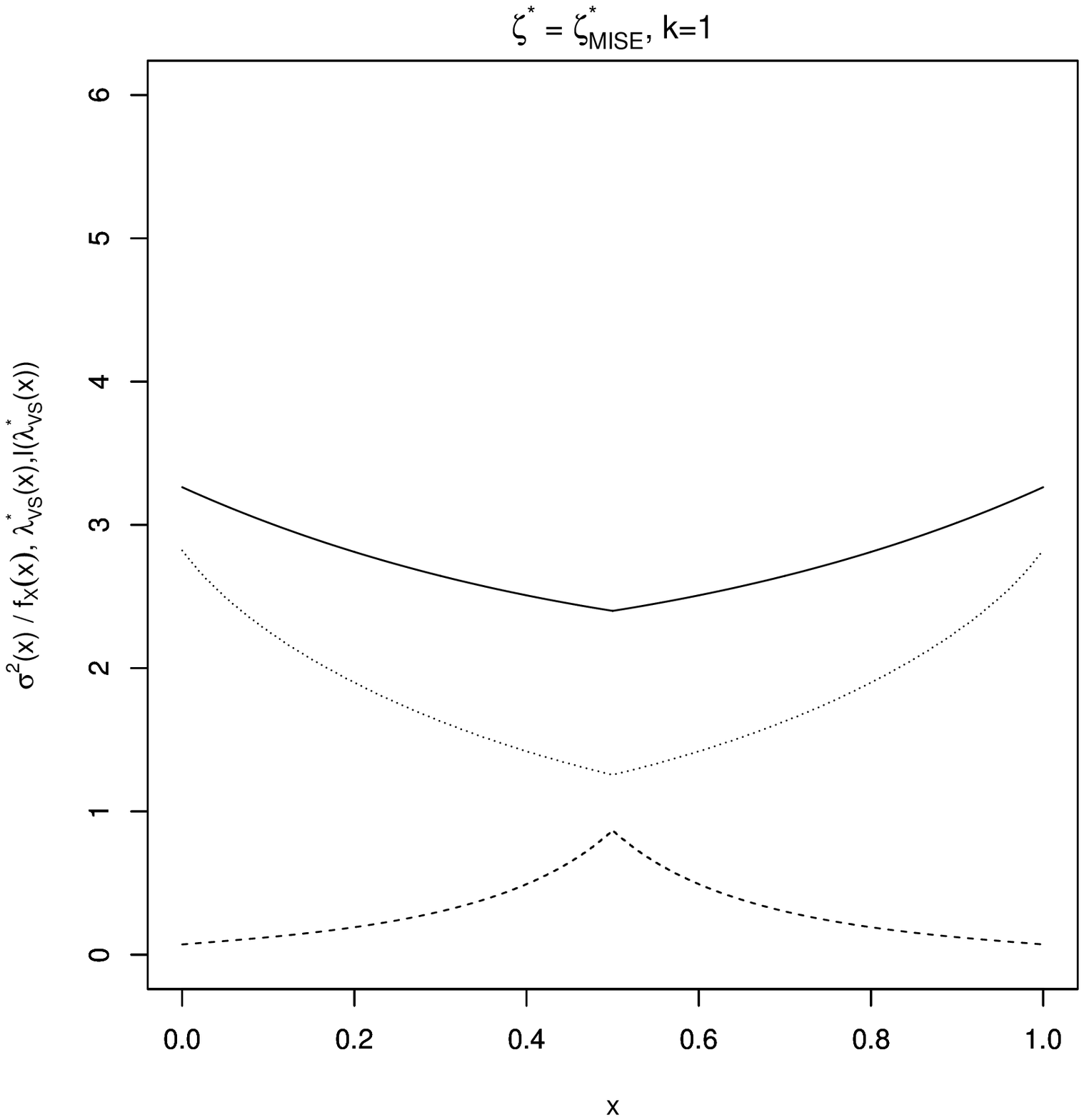}
\caption[]{Plots of $\sigma^{2}(x)/f_{X}(x)$, $l(\lambda_{VS}^{*}(x))$, and $\lambda_{VS}^{*}(x)$ at every $x$. $\zeta_{Var}^{*}$ and $\zeta_{MISE}^{*}$ are used in the left and right panels, respectively, with the settings used for Example~1-(i), $k=1$, and the Gaussian kernel.}
\label{Plot.VF.el.lambda}
\end{center}
\end{figure}

The SDs calculated in Step 7 can be regarded as degrees of variance stabilization. When running a simulation, we have to pay attention to the boundary effects discussed in Gasser and M\"uller (1979) and Rice (1984) for the fixed design kernel regression estimator. Boundary effects can occur even for the LL estimator in the areas $0 < x < h$ and $1-h < x < 1$, where the symmetry of the kernel is not satisfied and the estimator is biased. Especially, in the case of the CC estimator, the points $x-lh$ and $x+lh$ are outside the domain $I$ when $0 < x < l h$ and $1-lh < x < 1$, respectively, and the estimators are more biased around these areas. To avoid boundary effects, we set the domain as $[\iota, 1-\iota]$, $\iota = 0.0, 0.05, 0.10$ and $0.15$, and calculate the SD for every $\iota$. Tables \ref{bandwidth.and.AMISE.k1}, \ref{bandwidth.and.AMISE.k2}, and \ref{bandwidth.and.AMISE.k3} show the results of the simulation when $k=1, 2$, and $3$, respectively. In Figure \ref{Plot.VF.el.lambda}, we present the plots of $\gamma^{*}(x)$, $l(\lambda_{VS}^{*}(x))$, and $\lambda_{VS}^{*}(x)$ at every $x$ for both cases of $\zeta_{Var}^{*}$ and $\zeta_{MISE}^{*}$ used in the simulation.

First, we check which method performs the best in terms of the degree of variance stabilization by looking at the SD of the sample variances of the regression estimators calculated at an interval of 0.001 in the domain. Let us examine the cases $k=1, 2$, and $3$. The MSE-minimizing bandwidth yields the smallest SD in both groups [I] and [II] when $\iota = 0.0$, regardless of the sample size $n$. However, it is noticeable that, when $\iota = 0.10$, and $0.15$, either of the VS bandwidth or the VS weighting methods yields the smallest SD when compared with other estimators in each group, regardless of $n$. Notably, we find many cases in which the VS bandwidth method produces a smaller SD than the VS weighting method; however, it appears that the VS weighting method is superior to the VS bandwidth in terms of SD when the sample size is large $(n=1,000)$ and $\iota = 0.15$. These results obtained for $k=1,2$, and $3$ provide some evidence that the VS weighting method can stabilize estimator variance more efficiently than the VS bandwidth method in areas where boundary effects are ignored and the sample size is large.

Second, we examine the results of the simulation in terms of the AMISE and $\widehat{MISE}$. In this paper, we employ the strategy of stabilizing the estimator variance, rather than minimize the MISE or MSE. When using this strategy, it is inevitable that the VS bandwidth augments the MISE, compared with the MSE-minimizing bandwidth. However, in the case of the VS weighting method, it is expected that the MISE augmentation does not occur in all situations. In this sense, we are interested in calculating the cost of variance stabilization in terms of MISE augmentation. We present the AMISE values for each situation in Tables \ref{bandwidth.and.AMISE.k1}, \ref{bandwidth.and.AMISE.k2}, and \ref{bandwidth.and.AMISE.k3}. In group [II], the VS weighting method performs worse than the VS bandwidth in terms of the AMISE because the VS bandwidth uses two parameters $\lambda(x)$ and $h$ to reduce the AMISE, whereas the VS weighting method uses only one parameter $h$ for the same purpose. In group [I], the VS weighting method performs better than not only the VS bandwidth method but also the method using the MSE-minimizing bandwidth in terms of the AMISE. The strategy to minimize the estimator variance introduces the advantages of the VS weighting method in AMISE. The ratio of the AMISE calculated using the VS weighting method to that calculated using the MSE-minimizing bandwidth method takes the maximum value of $1.0667$ when $k=1$ in [II]. In the tables, we also present the $\widehat{MISE}$ values. The ratio of the $\widehat{MISE}$ estimated using the VS weighting method to that estimated using the MSE-minimizing bandwidth method takes the maximum value of $1.7156$ when $k=2$ and $n=1,000$ in [I].

Third, we compare $(a)$ and $(e)$, that is, $\lambda_{VS, Var}^{*}(x)$ and $\lambda_{VS, MISE}^{*}(x)$. We observe the estimator $(e)$ yields a larger $\widehat{MISE}$ for $k=1,2$, and $3$. This result appears odd, because $(e)$ would be expected to minimize MISE by definition. We interpret this result as being caused either by the remainder terms of the AMISE or by the fact that the estimator of the MISE in \eqref{estimator.MISE} requires improvement. If we fix the domain $\iota = 0.15$, the estimator $(e)$ yields a smaller SD for $k=1,2$, and $3$ regardless of the sample size $n$. If we emphasize variance stabilization, then the estimator $(e)$, the VS weighting method with $\lambda_{VS, MISE}^{*}(x)$, appears to be preferable. We observe the same tendency as to SD in the comparison between $(b)$ and $(f)$, i.e., between $\bar{\lambda}_{VS, Var}$ and $\bar{\lambda}_{VS, MISE}$.


Fourth, we explain the relationship of $\lambda_{VS}^{*}(x)$ and $l(\lambda_{VS}^{*}(x))$ with the $\gamma^{*}(x)$ expressed in Figure~\ref{Plot.VF.el.lambda} for the case $k=1$. The difference in calculating the two graphs in Figure~\ref{Plot.VF.el.lambda} is that the kernel and $\gamma^{*}(x)$ suffice to determine $\lambda_{VS, Var}^{*}(x)$, whereas additional information regarding $m^{(i)}(x), i=2,3$, and 4, is required to determine $\lambda_{VS, MISE}^{*}(x)$. Both graphs illustrate that the size of $\lambda_{VS}^{*}(x)$ is inversely proportional to that of the variance $\gamma^{*}(x)$ in achieving variance stabilization. In the domain in which the variance $\gamma^{*}(x)$ is low, the VS weighting method forces us to assign a higher weight for $\widehat{m_{h, \lambda}}(x \pm lh)$ to $\widehat{m_{h, \lambda}}(x)$ in the CC estimator and vice versa. This fact is compatible with the fact that $V(\lambda(x))$ increases with increasing $\lambda(x)$ exceeding a critical value $\lambda_{min}$, because the mechanism that reduces/increases the estimator variance in the domain where $\gamma^{*}(x)$ is high/low is required to achieve variance stabilization. For the cases $k=2$, and 3, we observe the same relationship.

\begin{figure}
\begin{center}
\includegraphics[width=0.425\textwidth]{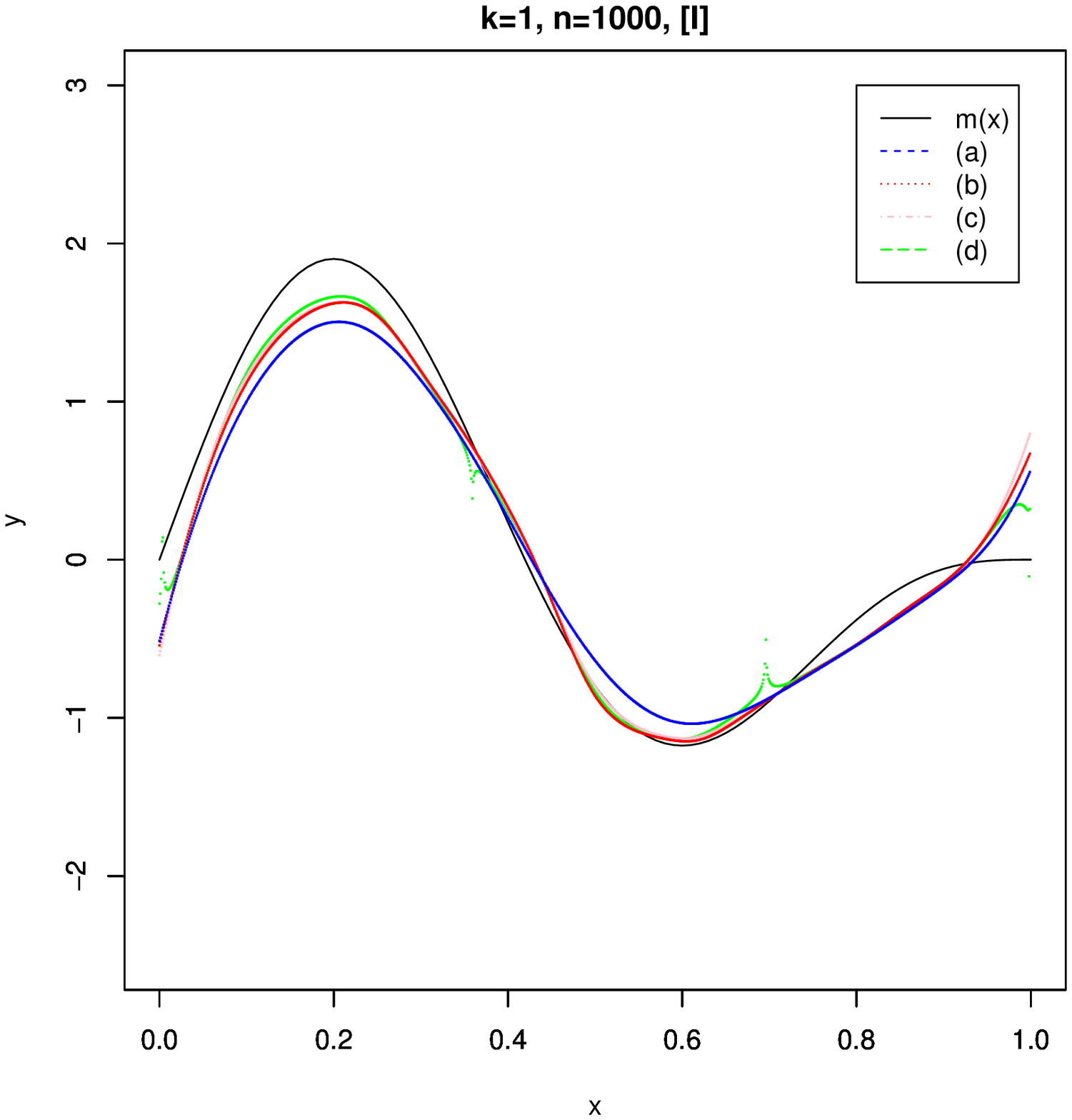}
\includegraphics[width=0.425\textwidth]{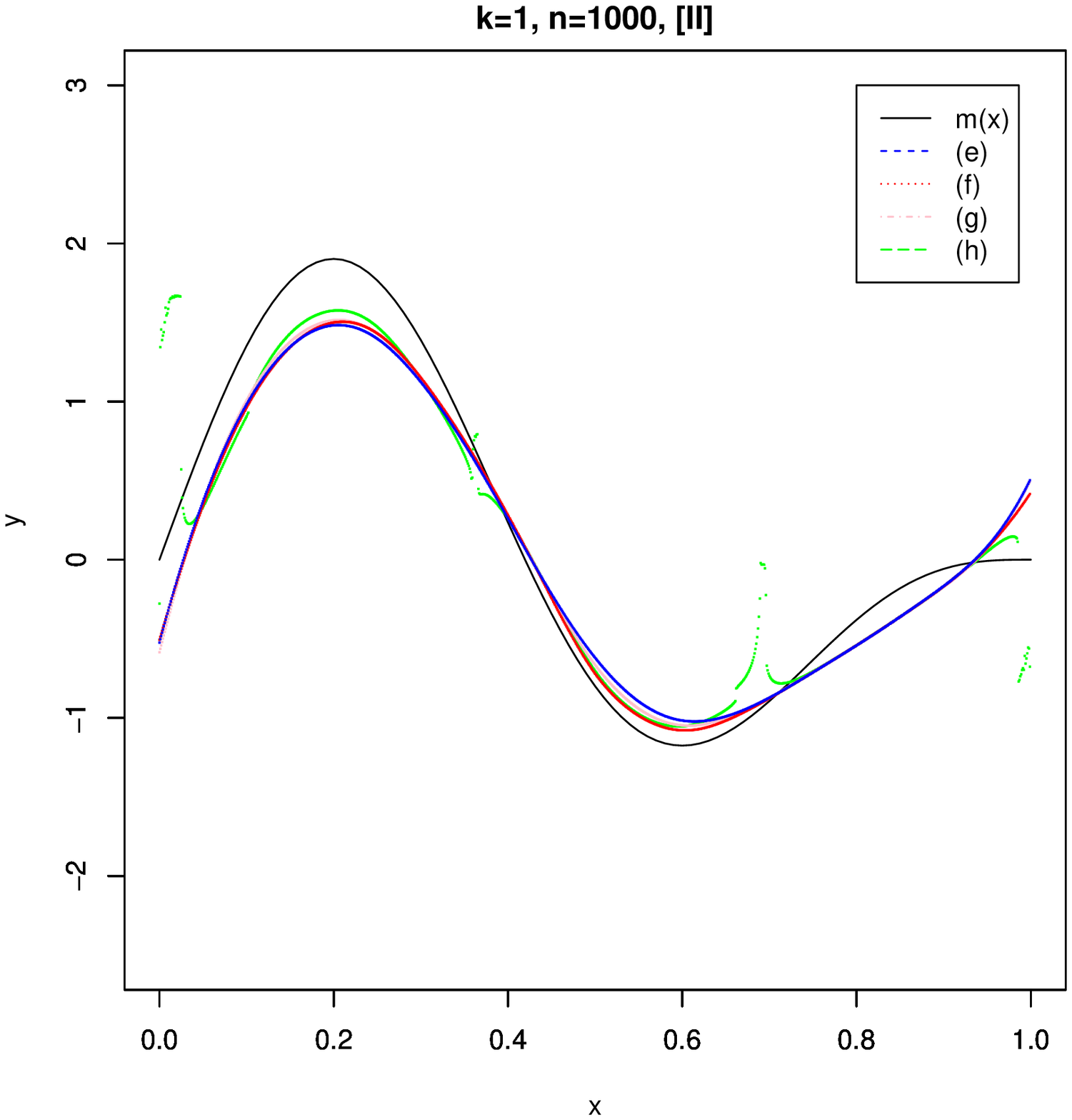}
\includegraphics[width=0.425\textwidth]{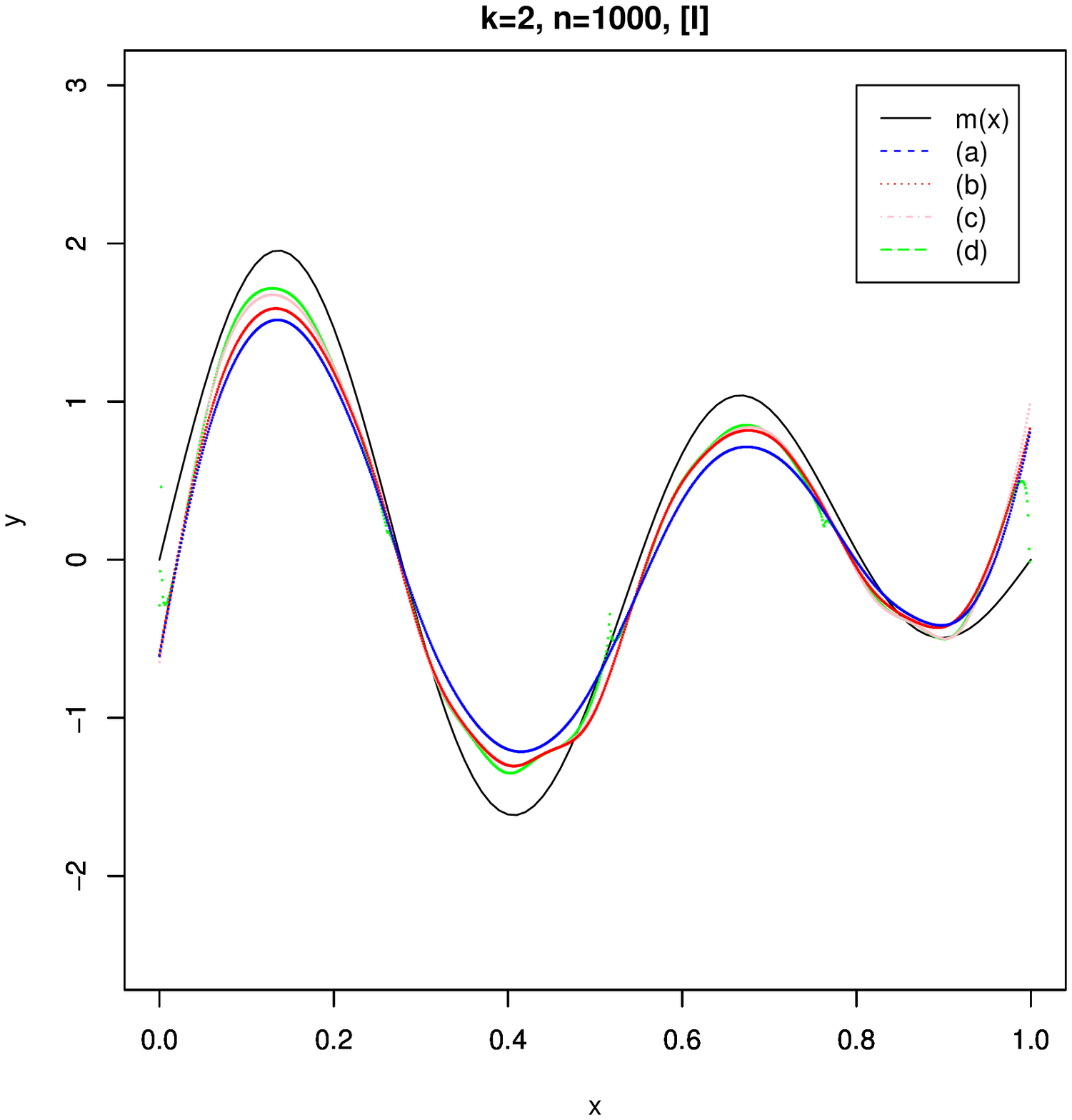}
\includegraphics[width=0.425\textwidth]{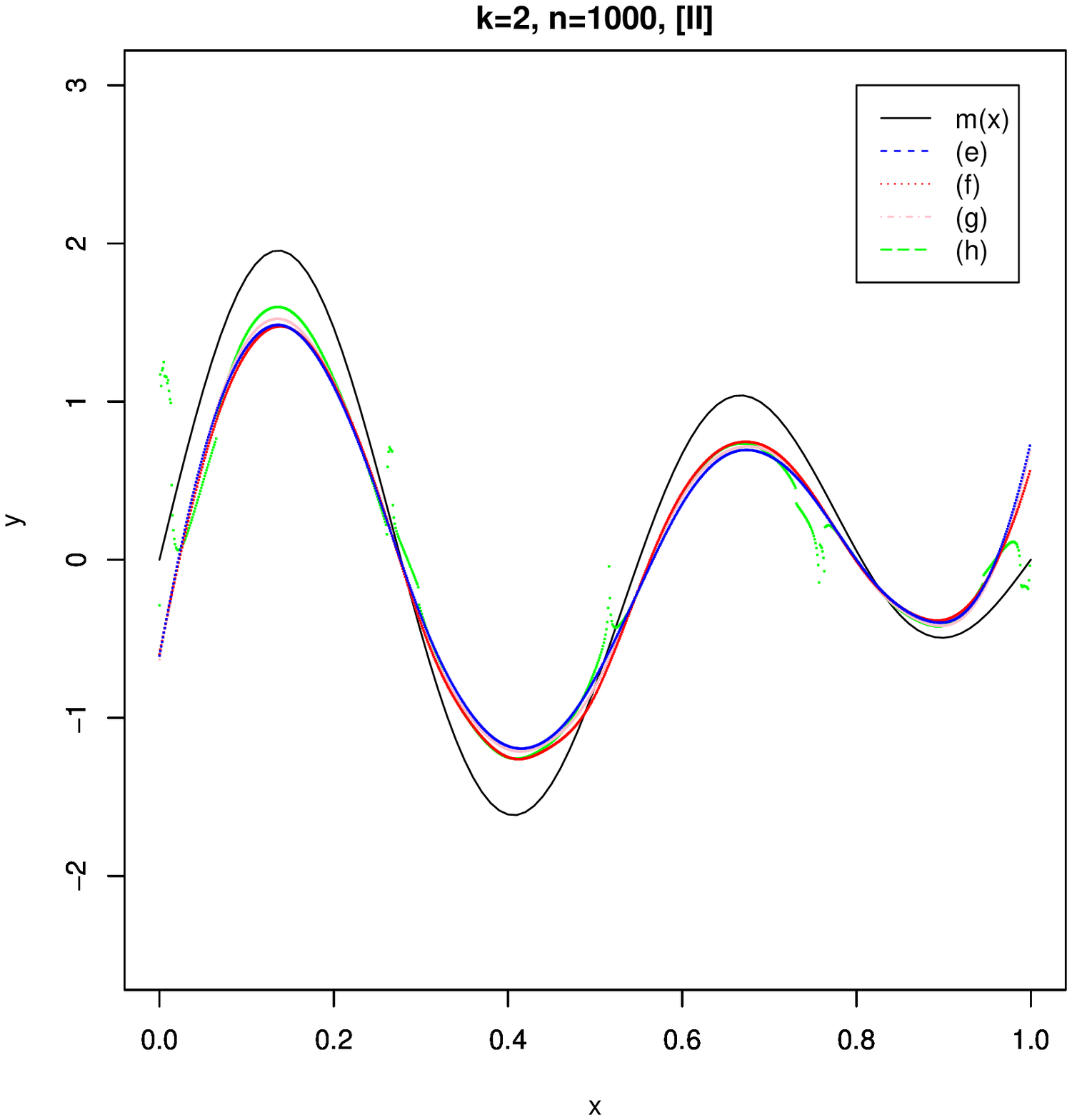}
\includegraphics[width=0.425\textwidth]{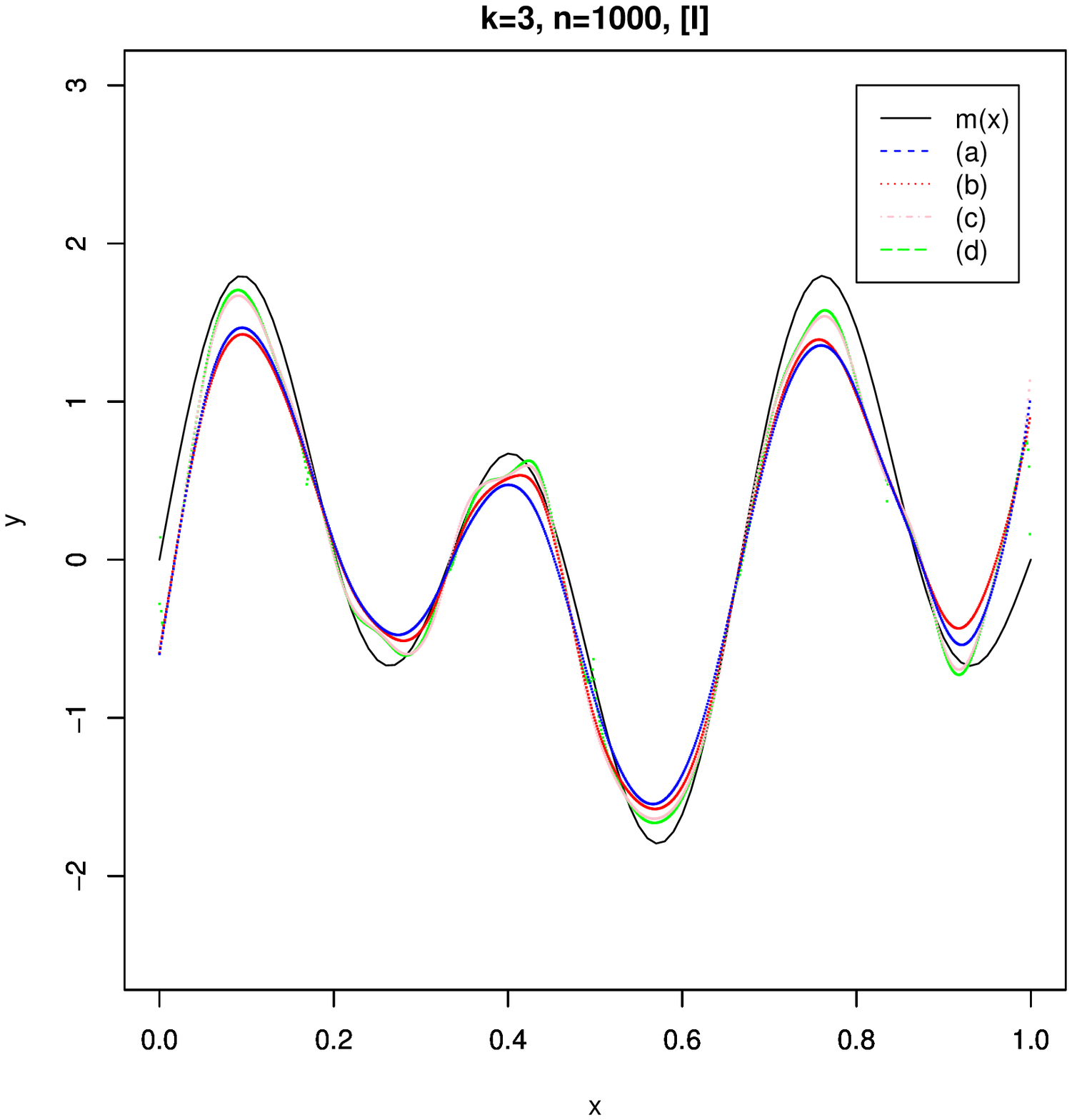}
\includegraphics[width=0.425\textwidth]{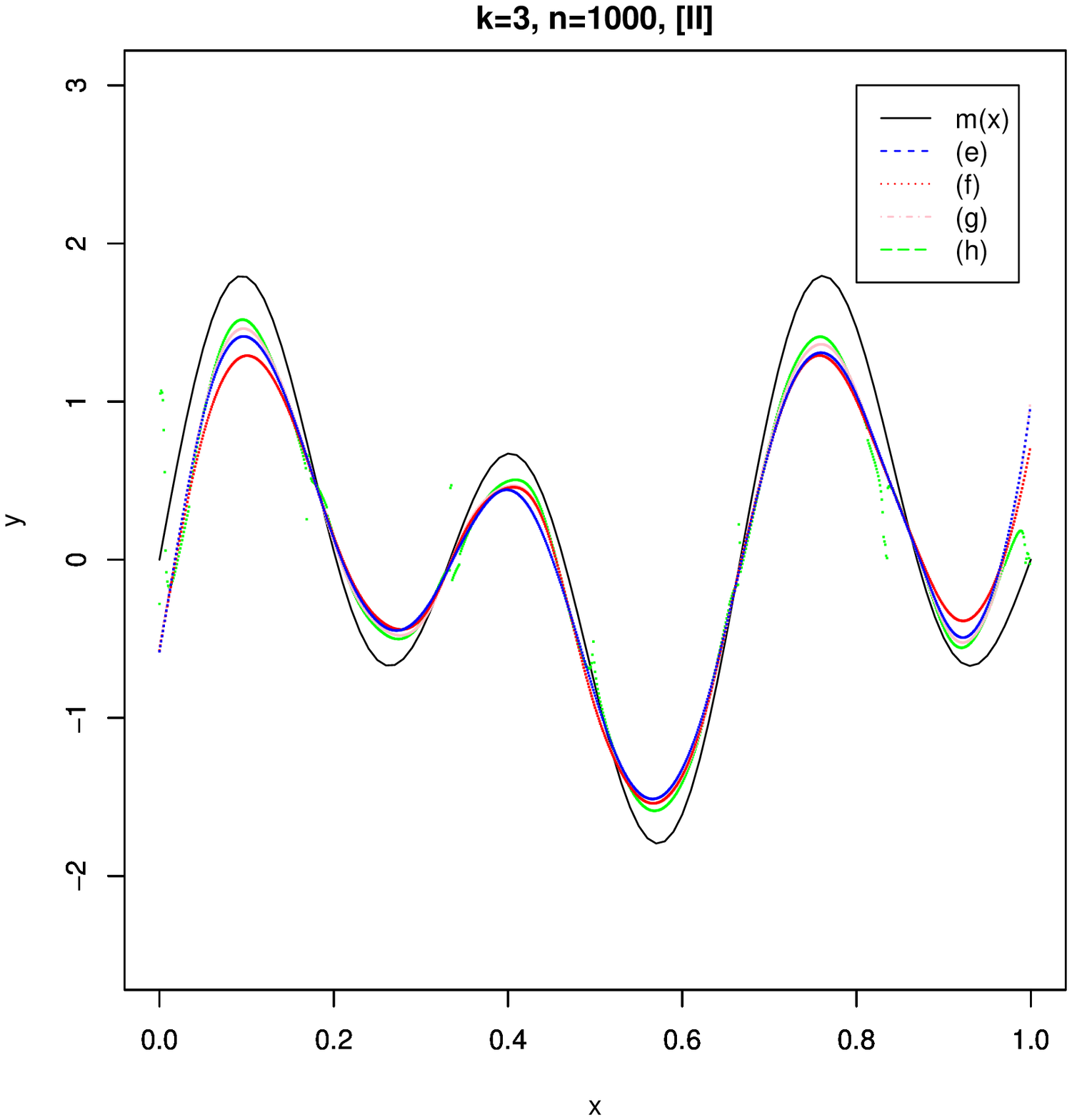}
\caption[]{Plots of the CC estimators $(a)$-$(h)$ for $k=1,2$, and $3$, and $n=1,000$. Black: true regression functions, blue: $(a)$$(e)$, red: $(b)$$(f)$, pink:$(c)$$(g)$, green:$(d)$$(h)$.}
\label{Regression.estimation}
\end{center}
\end{figure}

To assist in the visual understanding of the results, we also present the graphs of the eight estimators (shown in Figure \ref{Regression.estimation}) using a randomly selected sample of size $n=1,000$ for the cases $k=1,2$, and $3$. As pointed out by Nishida and Kanazawa (2015), we observe the discontinuous points in the curves of the regression estimators produced by the MSE-minimizing bandwidth $(h)$, which were produced more often than $(d)$, in the areas where $B(x)$ takes a zero value when $k=1$ and $2$ even though the curvature of $m(x)$ is not necessarily zero in those areas. These discontinuities in the regression estimator in $(h)$ causes the values of $\widehat{MISE}$ to be larger than the MISE values of other estimators even though $(h)$ is designed to minimize $\widehat{MISE}$. In our simulation, we attempt to employ the noisy setup for $\sigma^{2}(x)$ and $f_{X}(x)$ to emphasize the differences in variance stabilization between the estimators. Consequently, we observe that neither the conventional bandwidth method nor the two proposed methods work well in capturing the picture of the true regression functions around the peaks when $k=1,2$, and $3$. The costs of variance stabilization not only for the VS weighting method but also for the VS bandwidth method rise to the surface in the discrepancy between the true regression function and the regression estimator around the peaks.
\begin{landscape}
\begin{table}
\begin{center}
{\tiny{
\begin{tabular}{l|llll|lllll}
\hline
\hline
$k=1$ & [I] & & & & [II] & & & & \\
$M=100$& ${h}_{VS, Var}^{*}$ & $h_{VS, Var}(x)$ & ${h}_{fixed, Var}$ & $h_{MSE, Var}(x)$ & ${h}_{VS, MISE}^{*}$ & $h_{VS, MISE}(x)$ & ${h}_{fixed, MISE}$ & $h_{MSE, MISE}(x)$ & \\
& $\lambda_{VS, Var}^{*}(x)$ & $\bar{\lambda}_{VS, Var}$ & $\bar{\lambda}_{fixed, Var}$ & $\bar{\lambda}_{MSE, Var}$ & $\lambda_{VS, MISE}^{*}(x)$ & $\bar{\lambda}_{VS, MISE}$ & $\bar{\lambda}_{fixed, MISE}$ & $\lambda_{MSE, MISE}(x)$ \\
 & $(a)$ & $(b)$ & $(c)$ & $(d)$ & $(e)$ & $(f)$ & $(g)$ & $(h)$ \\
\hline
$n=100$ & & & & & & & & & \\
Bandwidth \ \ & 0.0791 & 0.0218  & 0.0606 & variable & 0.0818 & 0.0283 & 0.0777 & variable \\
${AMISE}$ & 1.2161 $\cdot 10^{-1}$ & 1.3272 $\cdot 10^{-1}$ & 1.3168 $\cdot 10^{-1}$ & 1.2236 $\cdot 10^{-1}$ & 1.2058 $\cdot 10^{-1}$ & 1.2011 $\cdot 10^{-1}$ & 1.1940 $\cdot 10^{-1}$ & 1.1303 $\cdot 10^{-1}$ \\
$\widehat{MISE}$ & 2.4415 $\cdot 10^{-1}$ & 2.3604 $\cdot 10^{-1}$ & 2.4182 $\cdot 10^{-1}$ & 2.1398 $\cdot 10^{-1}$ & 2.5019 $\cdot 10^{-1}$ & 2.3449 $\cdot 10^{-1}$ & 2.4262 $\cdot 10^{-1}$ & 2.6151 $\cdot 10^{-1}$ \\
$\lambda$ & variable & 0.0376 & 0.0376 & 0.0376  & variable & 0.158 & 0.147 & variable \\
$\zeta^{*}$ & 0.8413 & --- & --- & --- & 0.8774 & --- & --- & --- \\
SD$(\iota=0.00)$ & 2.0263 $\cdot 10^{-1}$ & 2.1268 $\cdot 10^{-1}$ & 2.5362 $\cdot 10^{-1}$ & 1.5309 $\cdot 10^{-1}$ & 1.9674 $\cdot 10^{-1}$ & 1.7589 $\cdot 10^{-1}$ & 2.0734 $\cdot 10^{-1}$ & 9.3426 $\cdot 10^{-2}$ \\
\ \ \ \ $(\iota=0.05)$ & 6.3499 $\cdot 10^{-2}$ & 5.4056 $\cdot 10^{-2}$ & 5.5982 $\cdot 10^{-2}$ & 6.0395 $\cdot 10^{-2}$ & 6.4098 $\cdot 10^{-2}$ & 6.0434 $\cdot 10^{-2}$ & 6.1260 $\cdot 10^{-2}$ & 6.3026 $\cdot 10^{-2}$ \\
\ \ \ \ $(\iota=0.10)$ & 1.9597 $\cdot 10^{-2}$ & 1.6050 $\cdot 10^{-2}$ & 1.6812 $\cdot 10^{-2}$ & 2.6897$\cdot 10^{-2}$ & 2.0407 $\cdot 10^{-2}$ & 1.7495	$\cdot 10^{-2}$ & 1.7844 $\cdot 10^{-2}$ & 2.3144 $\cdot 10^{-2}$ \\
\ \ \ \ $(\iota=0.15)$ & 8.8733 $\cdot 10^{-3}$ & 1.4477 $\cdot 10^{-2}$ & 1.3340 $\cdot 10^{-2}$ & 2.5086 $\cdot 10^{-2}$ & 8.7339 $\cdot 10^{-3}$ & 8.0422 $\cdot 10^{-3}$ & 8.2292 $\cdot 10^{-3}$ & 1.6542 $\cdot 10^{-2}$ \\
\hline
$n=500$ & & & & & & & & & \\
Bandwidth \ \ & 0.0661 & 0.0182 & 0.0507 & variable & 0.0684 & 0.0236 & 0.0650 & variable \\
${AMISE}$ & 2.9084 $\cdot 10^{-2}$ & 3.1743 $\cdot 10^{-2}$ & 3.1493 $\cdot 10^{-2}$ & 2.9265 $\cdot 10^{-2}$ & 2.8839 $\cdot 10^{-2}$ & 2.8725 $\cdot 10^{-2}$ & 2.8556 $\cdot 10^{-2}$ & 2.7034 $\cdot 10^{-2}$ \\
$\widehat{MISE}$ & 7.8654 $\cdot 10^{-2}$ & 5.8929 $\cdot 10^{-2}$ & 5.9722 $\cdot 10^{-2}$ & 5.4875$\cdot 10^{-2}$ & 8.415 $\cdot 10^{-2}$ & 7.4702 $\cdot 10^{-2}$ & 7.6491 $\cdot 10^{-2}$ & 1.2578$\cdot 10^{-1}$ \\
$\lambda$ & variable & 0.0376 & 0.0376 & 0.0376	& variable & 0.158 & 0.147 & variable \\
$\zeta^{*}$ & 0.8413 & --- & --- & --- & 0.8774 & --- & --- & --- \\
SD$(\iota=0.00)$ & 4.6670 $\cdot 10^{-2}$ & 4.7025 $\cdot 10^{-2}$ & 5.2344 $\cdot 10^{-2}$ & 3.5919 $\cdot 10^{-2}$ & 4.5840 $\cdot 10^{-2}$ & 4.2065 $\cdot 10^{-2}$ & 4.7236 $\cdot 10^{-2}$ & 2.3363 $\cdot 10^{-2}$ \\
\ \ \ \ $(\iota=0.05)$ & 1.3674 $\cdot 10^{-2}$ & 1.1315 $\cdot 10^{-2}$ & 1.1737 $\cdot 10^{-2}$ & 1.2581$\cdot 10^{-2}$ & 1.3885 $\cdot 10^{-2}$ & 1.3151 $\cdot 10^{-2}$ & 1.3066 $\cdot 10^{-2}$ & 1.4734 $\cdot 10^{-2}$ \\
\ \ \ \ $(\iota=0.10)$ & 5.0507 $\cdot 10^{-3}$ & 4.0034 $\cdot 10^{-3}$ & 4.7618 $\cdot 10^{-3}$ & 5.7988$\cdot 10^{-3}$ & 5.1448 $\cdot 10^{-3}$ & 4.2162 $\cdot 10^{-3}$ & 4.6357 $\cdot 10^{-3}$ & 5.4762 $\cdot 10^{-3}$ \\
\ \ \ \ $(\iota=0.15)$ & 2.9274 $\cdot 10^{-3}$ & 3.2524 $\cdot 10^{-3}$ & 3.5920 $\cdot 10^{-3}$ & 4.5805 $\cdot 10^{-3}$ & 2.8671 $\cdot 10^{-3}$ & 2.2670 $\cdot 10^{-3}$ & 2.7512 $\cdot 10^{-3}$ & 3.7204 $\cdot 10^{-3}$ \\
\hline
$n=1,000$ & & & & & & & & & \\
Bandwidth \ \ & 0.0612 & 0.0169 & 0.0469 & variable & 0.0633 & 0.0219 & 0.0601 & variable \\
${AMISE}$ & 1.5706 $\cdot 10^{-2}$ & 1.7142 $\cdot 10^{-2}$ & 1.7007 $\cdot 10^{-2}$ & 1.5804$\cdot 10^{-2}$ & 1.5574 $\cdot 10^{-2}$ & 1.5512 $\cdot 10^{-2}$ & 1.5421 $\cdot 10^{-2}$ & 1.4599 $\cdot 10^{-2}$ \\
$\widehat{MISE}$ & 5.3153 $\cdot 10^{-2}$ & 3.3869 $\cdot 10^{-2}$ & 3.4454 $\cdot 10^{-2}$ & 3.2153 $\cdot 10^{-2}$ & 5.8104 $\cdot 10^{-2}$ & 5.0623 $\cdot 10^{-2}$ & 5.1593 $\cdot 10^{-2}$ & 1.0009 $\cdot 10^{-1}$ \\
$\lambda$ & variable & 0.0376 & 0.0376 & 0.0376 & variable & 0.158 & 0.147 & variable \\
$\zeta^{*}$ & 0.8413 & --- & --- & --- & 0.8774 & --- & --- & --- \\
SD$(\iota=0.00)$ & 2.3465 $\cdot 10^{-2}$ & 2.3487 $\cdot 10^{-2}$ & 2.6742 $\cdot 10^{-2}$ & 1.6294 $\cdot 10^{-2}$ & 2.3309 $\cdot 10^{-2}$ & 2.1794	$\cdot 10^{-2}$ & 2.4818 $\cdot 10^{-2}$ & 1.0044 $\cdot 10^{-2}$ \\
\ \ \ \ $(\iota=0.05)$ & 4.6555 $\cdot 10^{-3}$ & 3.3048 $\cdot 10^{-3}$ & 3.1998 $\cdot 10^{-3}$ & 4.3583 $\cdot 10^{-3}$ & 4.8455 $\cdot 10^{-3}$ & 4.7449 $\cdot 10^{-3}$ & 4.5089 $\cdot 10^{-3}$ & 5.6071 $\cdot 10^{-3}$ \\
\ \ \ \ $(\iota=0.10)$ & 1.5848 $\cdot 10^{-3}$ & 1.8519 $\cdot 10^{-3}$ & 1.8445 $\cdot 10^{-3}$ & 3.2496 $\cdot 10^{-3}$ & 1.5588 $\cdot 10^{-3}$ & 1.4512 $\cdot 10^{-3}$ & 1.4942 $\cdot 10^{-3}$ & 2.4198 $\cdot 10^{-3}$ \\
\ \ \ \ $(\iota=0.15)$ & 1.1925 $\cdot 10^{-3}$ & 1.7253 $\cdot 10^{-3}$ & 1.6142 $\cdot 10^{-3}$ & 3.0974 $\cdot 10^{-3}$  & 1.1514 $\cdot 10^{-3}$ & 1.2440 $\cdot 10^{-3}$ & 1.1814 $\cdot 10^{-3}$ & 2.2798 $\cdot 10^{-3}$ \\
\hline
\hline
\end{tabular}
\caption[]
{Comparison of the eight estimators. We employ $m(x) = m_{k}(x) = \frac{2}{5} \left[ 3 \sin(2k\pi x) + 2 \sin(3 \pi x) \right]$, $k=1$, $f_{X}(x) = (1/0.3829)(1/\sqrt{2\pi}) \exp \left(-0.5(x-0.5)^2 \right)$, $\sigma^2(x) = (2.5 + |x - 0.5|)$ defined in $I=[0, 1]$ and Gaussian kernel.} \label{bandwidth.and.AMISE.k1}
}}
\end{center}
\end{table}
\end{landscape}
\begin{landscape}
\begin{table}
\begin{center}
{\tiny{
\begin{tabular}{l|llll|lllll}
\hline
\hline
$k=2$ & [I] & & & & [II] & & & & \\
$M=100$& ${h}_{VS, Var}^{*}$ & $h_{VS, Var}(x)$ & ${h}_{fixed, Var}$ & $h_{MSE, Var}(x)$ & ${h}_{VS, MISE}^{*}$ & $h_{VS, MISE}(x)$ & ${h}_{fixed, MISE}$ & $h_{MSE, MISE}(x)$ & \\
& $\lambda_{VS, Var}^{*}(x)$ & $\bar{\lambda}_{VS, Var}$ & $\bar{\lambda}_{fixed, Var}$ & $\bar{\lambda}_{MSE, Var}$ & $\lambda_{VS, MISE}^{*}(x)$ & $\bar{\lambda}_{VS, MISE}$ & $\bar{\lambda}_{fixed, MISE}$ & $\lambda_{MSE, MISE}(x)$ \\
 & $(a)$ & $(b)$ & $(c)$ & $(d)$ & $(e)$ & $(f)$ & $(g)$ & $(h)$ \\
\hline
$n=100$ & & & & & & & & & \\
Bandwidth \ \ & 0.0558 & 0.0173 & 0.0431 & variable & 0.0581 & 0.0201 & 0.0553 & variable \\
${AMISE}$ & 1.7219 $\cdot 10^{-1}$ & 1.8766 $\cdot 10^{-1}$ & 1.8516 $\cdot 10^{-1}$ & 1.7312 $\cdot 10^{-1}$ & 1.7054 $\cdot 10^{-1}$ & 1.6994 $\cdot 10^{-1}$ & 1.6786 $\cdot 10^{-1}$ & 1.6021 $\cdot 10^{-1}$ \\
$\widehat{MISE}$ & 3.2369 $\cdot 10^{-1}$ & 3.0355 $\cdot 10^{-1}$ & 3.1772 $\cdot 10^{-1}$ & 2.7610 $\cdot 10^{-1}$ & 3.3537 $\cdot 10^{-1}$ & 3.1368 $\cdot 10^{-1}$ & 3.2452 $\cdot 10^{-1}$ & 2.8374 $\cdot 10^{-1}$ \\
$\lambda$ & variable & 0.0376 & 0.0376 & 0.0376 & variable & 0.154 & 0.148 & variable \\
$\zeta^{*}$ & 0.8413 & --- & --- & --- & 0.8814 & --- & --- & --- \\
SD$(\iota=0.00)$ & 2.8370 $\cdot 10^{-1}$ & 2.6590 $\cdot 10^{-1}$ & 3.4851 $\cdot 10^{-1}$ & 2.0942 $\cdot 10^{-1}$ & 2.7419 $\cdot 10^{-1}$ & 2.4044 $\cdot 10^{-1}$ & 2.9161 $\cdot 10^{-1}$ & 1.2544 $\cdot 10^{-1}$ \\
\ \ \ \ $(\iota=0.05)$ & 5.2597 $\cdot 10^{-2}$ & 4.7541 $\cdot 10^{-2}$ & 4.7920 $\cdot 10^{-2}$ & 5.5785 $\cdot 10^{-2}$ & 5.3461 $\cdot 10^{-2}$ & 4.9680 $\cdot 10^{-2}$ & 4.9998 $\cdot 10^{-2}$ & 5.7225 $\cdot 10^{-2}$ \\
\ \ \ \ $(\iota=0.10)$ & 1.6709 $\cdot 10^{-2}$ & 2.2964 $\cdot 10^{-2}$ & 2.6658 $\cdot 10^{-2}$ & 3.5898 $\cdot 10^{-2}$ & 1.6026 $\cdot 10^{-2}$ & 1.5348 $\cdot 10^{-2}$ &  1.5533 $\cdot 10^{-2}$ & 2.5415 $\cdot 10^{-2}$ \\
\ \ \ \ $(\iota=0.15)$ & 1.4219 $\cdot 10^{-2}$ & 2.3855 $\cdot 10^{-2}$ & 2.6750 $\cdot 10^{-2}$ & 3.3891 $\cdot 10^{-2}$ & 1.2908 $\cdot 10^{-2}$ & 1.5155 $\cdot 10^{-2}$ & 1.4257 $\cdot 10^{-2}$ & 2.4387 $\cdot 10^{-2}$ \\
\hline
$n=500$ & & & & & & & & & \\
Bandwidth \ \ & 0.0464 & 0.0144 & 0.0360 & variable & 0.0486 & 0.0168 & 0.0463 & variable \\
${AMISE}$ & 4.1183 $\cdot 10^{-2}$ & 4.4882 $\cdot 10^{-2}$ & 4.4284 $\cdot 10^{-2}$ & 4.1403 $\cdot 10^{-2}$ & 4.0788 $\cdot 10^{-2}$ & 4.0643 $\cdot 10^{-2}$ & 4.0148 $\cdot 10^{-2}$ & 3.8316 $\cdot 10^{-2}$ \\
$\widehat{MISE}$ & 1.0744 $\cdot 10^{-1}$ & 8.4192 $\cdot 10^{-2}$ & 7.9577 $\cdot 10^{-2}$ & 7.3164 $\cdot 10^{-2}$ & 1.1729 $\cdot 10^{-1}$ & 1.0429 $\cdot 10^{-1}$ & 1.0615 $\cdot 10^{-1}$ & 1.1164 $\cdot 10^{-1}$ \\
$\lambda$ & variable & 0.0376 &	0.0376 & 0.0376 & variable & 0.154 & 0.148 & variable \\
$\zeta^{*}$ & 0.8413 & --- & --- & --- & 0.8814 & --- & --- & --- \\
SD$(\iota=0.00)$ & 5.6026 $\cdot 10^{-2}$ & 5.2671 $\cdot 10^{-2}$ & 6.2085 $\cdot 10^{-2}$ & 4.3207 $\cdot 10^{-2}$ & 5.5157 $\cdot 10^{-2}$ & 5.0985 $\cdot 10^{-2}$ & 5.7117 $\cdot 10^{-2}$ & 2.9233 $\cdot 10^{-2}$ \\
\ \ \ \ $(\iota=0.05)$ & 1.1008 $\cdot 10^{-2}$ & 9.3196 $\cdot 10^{-3}$ & 9.4923 $\cdot 10^{-3}$ & 1.1793 $\cdot 10^{-2}$ & 1.1175 $\cdot 10^{-2}$ & 1.0044 $\cdot 10^{-2}$ & 1.0117 $\cdot 10^{-2}$ & 1.1894 $\cdot 10^{-2}$ \\
\ \ \ \ $(\iota=0.10)$ & 4.7684 $\cdot 10^{-3}$ & 4.0196 $\cdot 10^{-3}$ & 5.2272 $\cdot 10^{-3}$ & 8.4053 $\cdot 10^{-3}$ & 4.6449 $\cdot 10^{-3}$ & 3.5150 $\cdot 10^{-3}$ & 4.3186 $\cdot 10^{-3}$ & 6.4205 $\cdot 10^{-3}$ \\
\ \ \ \ $(\iota=0.15)$ & 3.6027 $\cdot 10^{-3}$ & 3.7480 $\cdot 10^{-3}$ &	4.5930 $\cdot 10^{-3}$ & 7.8751 $\cdot 10^{-3}$ & 3.4174 $\cdot 10^{-3}$ & 2.9820 $\cdot 10^{-3}$ & 3.3970 $\cdot 10^{-3}$ & 5.6201 $\cdot 10^{-3}$ \\
\hline
$n=1,000$ & & & & & & & & & \\
Bandwidth \ \ & 0.0432 & 0.0134 & 0.0334 & variable & 0.0450 & 0.0155 & 0.0428 & variable \\
${AMISE}$ & 2.224 $\cdot 10^{-2}$ & 2.4237 $\cdot 10^{-2}$ & 2.3915 $\cdot 10^{-2}$ & 2.2359 $\cdot 10^{-2}$ & 2.2027 $\cdot 10^{-2}$ & 2.1948 $\cdot 10^{-2}$ & 2.1681 $\cdot 10^{-2}$ & 2.0691 $\cdot 10^{-2}$ \\
$\widehat{MISE}$ & 6.9514 $\cdot 10^{-2}$ & 4.961 $\cdot 10^{-2}$ & 4.4544 $\cdot 10^{-2}$ & 4.0517 $\cdot 10^{-2}$ & 7.7784 $\cdot 10^{-2}$ & 6.8063 $\cdot 10^{-2}$ & 6.8699 $\cdot 10^{-2}$ & 7.8068 $\cdot 10^{-2}$ \\
$\lambda$ & variable & 0.0376 & 0.0376 & 0.0376 & variable & 0.154 & 0.148 & variable \\
$\zeta^{*}$ & 0.8413 & --- & --- & --- & 0.8814 & --- & --- & --- \\
SD$(\iota=0.00)$ & 2.9469 $\cdot 10^{-2}$ & 2.7383 $\cdot 10^{-2}$ & 3.2711 $\cdot 10^{-2}$ & 2.0869 $\cdot 10^{-2}$ & 2.9219 $\cdot 10^{-2}$ & 2.7316 $\cdot 10^{-2}$ & 3.0795 $\cdot 10^{-2}$ & 1.4105 $\cdot 10^{-2}$ \\
\ \ \ \ $(\iota=0.05)$ & 3.1505 $\cdot 10^{-3}$ & 2.7050 $\cdot 10^{-3}$ & 2.7023 $\cdot 10^{-3}$ & 4.2393 $\cdot 10^{-3}$ & 3.2990 $\cdot 10^{-3}$ & 3.1903 $\cdot 10^{-3}$ & 2.9771 $\cdot 10^{-3}$ & 4.6211 $\cdot 10^{-3}$ \\
\ \ \ \ $(\iota=0.10)$ & 2.0067 $\cdot 10^{-3}$ & 2.3463 $\cdot 10^{-3}$ & 2.6463 $\cdot 10^{-3}$ & 4.2181 $\cdot 10^{-3}$ & 1.8964 $\cdot 10^{-3}$ & 1.8560 $\cdot 10^{-3}$ & 1.9006 $\cdot 10^{-3}$ & 3.3040 $\cdot 10^{-3}$ \\
\ \ \ \ $(\iota=0.15)$ & 1.6092 $\cdot 10^{-3}$ & 2.1438 $\cdot 10^{-3}$ & 2.4511 $\cdot 10^{-3}$ & 4.0180 $\cdot 10^{-3}$ & 1.5035 $\cdot 10^{-3}$ & 1.5701 $\cdot 10^{-3}$ & 1.5402 $\cdot 10^{-3}$ & 3.0453 $\cdot 10^{-3}$ \\
\hline
\hline
\end{tabular}
\caption[]
{Comparison of the eight estimators. We employ $m(x) = m_{k}(x) = \frac{2}{5} \left[ 3 \sin(2k\pi x) + 2 \sin(3 \pi x) \right]$, $k=2$, $f_{X}(x) = (1/0.3829)(1/\sqrt{2\pi}) \exp \left(-0.5(x-0.5)^2 \right)$, $\sigma^2(x) = (2.5 + |x - 0.5|)$ defined in $I=[0, 1]$ and Gaussian kernel.} \label{bandwidth.and.AMISE.k2}
}}
\end{center}
\end{table}
\end{landscape}
\begin{landscape}
\begin{table}
\begin{center}
{\tiny{
\begin{tabular}{l|llll|lllll}
\hline
\hline
$k=3$ & [I] & & & & [II] & & & & \\
$M=100$& ${h}_{VS, Var}^{*}$ & $h_{VS, Var}(x)$ & ${h}_{fixed, Var}$ & $h_{MSE, Var}(x)$ & ${h}_{VS, MISE}^{*}$ & $h_{VS, MISE}(x)$ & ${h}_{fixed, MISE}$ & $h_{MSE, MISE}(x)$ & \\
& $\lambda_{VS, Var}^{*}(x)$ & $\bar{\lambda}_{VS, Var}$ & $\bar{\lambda}_{fixed, Var}$ & $\bar{\lambda}_{MSE, Var}$ & $\lambda_{VS, MISE}^{*}(x)$ & $\bar{\lambda}_{VS, MISE}$ & $\bar{\lambda}_{fixed, MISE}$ & $\lambda_{MSE, MISE}(x)$ \\
 & $(a)$ & $(b)$ & $(c)$ & $(d)$ & $(e)$ & $(f)$ & $(g)$ & $(h)$ \\
\hline
$n=100$ & & & & & & & & & \\
Bandwidth \ \ & 0.0389 & 0.0147 & 0.0302 & variable & 0.0412 & 0.0158 & 0.0387 & variable \\
${AMISE}$ & 2.4698 $\cdot 10^{-1}$ & 2.6858 $\cdot 10^{-1}$ & 2.6420 $\cdot 10^{-1}$ & 2.4833 $\cdot 10^{-1}$ & 2.4417 $\cdot 10^{-1}$ & 2.4333 $\cdot 10^{-1}$ & 2.3950 $\cdot 10^{-1}$ & 2.3014 $\cdot 10^{-1}$ \\
$\widehat{MISE}$ & 3.9078 $\cdot 10^{-1}$ & 3.7227 $\cdot 10^{-1}$ & 4.1538 $\cdot 10^{-1}$ & 3.6652 $\cdot 10^{-1}$ & 4.0490 $\cdot 10^{-1}$ & 3.9884 $\cdot 10^{-1}$ & 3.9461 $\cdot 10^{-1}$ & 3.4084 $\cdot 10^{-1}$ \\
$\lambda$ & variable & 0.0376 & 0.0376 & 0.0376 & variable & 0.151 & 0.148 & variable \\
$\zeta^{*}$ & 0.8413 & --- & --- & --- & 0.8954 & --- & --- & --- \\
SD$(\iota=0.00)$ & 4.0456 $\cdot 10^{-1}$ & 3.1283 $\cdot 10^{-1}$ & 5.1397 $\cdot 10^{-1}$ & 2.9982 $\cdot 10^{-1}$ & 3.9130 $\cdot 10^{-1}$ & 3.1081 $\cdot 10^{-1}$ & 4.2389 $\cdot 10^{-1}$ & 1.9036 $\cdot 10^{-1}$ \\
\ \ \ \ $(\iota=0.05)$ & 4.3068 $\cdot 10^{-2}$ & 4.4562 $\cdot 10^{-2}$ & 4.9678 $\cdot 10^{-2}$ & 7.0954 $\cdot 10^{-2}$ & 4.2768 $\cdot 10^{-2}$ & 4.2200 $\cdot 10^{-2}$ & 3.9168 $\cdot 10^{-2}$ & 5.0725 $\cdot 10^{-2}$ \\
\ \ \ \ $(\iota=0.10)$ & 2.3219 $\cdot 10^{-2}$ & 2.7677 $\cdot 10^{-2}$ & 4.4088 $\cdot 10^{-2}$ & 6.6200 $\cdot 10^{-2}$ & 1.9342 $\cdot 10^{-2}$ & 1.8783 $\cdot 10^{-2}$ & 2.1625 $\cdot 10^{-2}$ & 3.7747 $\cdot 10^{-2}$ \\
\ \ \ \ $(\iota=0.15)$ & 2.1988 $\cdot 10^{-2}$ & 2.8752 $\cdot 10^{-2}$ & 4.3407 $\cdot 10^{-2}$ & 6.6047 $\cdot 10^{-2}$ & 1.8405 $\cdot 10^{-2}$ & 1.9579 $\cdot 10^{-2}$ & 2.1777 $\cdot 10^{-2}$ & 3.7987 $\cdot 10^{-2}$ \\
\hline
$n=500$ & & & & & & & & & \\
Bandwidth \ \ & 0.0325 & 0.0122 & 0.0252 & variable & 0.0345 & 0.0132 & 0.0324 & variable \\
${AMISE}$ & 5.907 $\cdot 10^{-2}$ & 6.4234 $\cdot 10^{-2}$ & 6.3188 $\cdot 10^{-2}$ & 5.9392 $\cdot 10^{-2}$ & 5.8397 $\cdot 10^{-2}$ & 5.8197 $\cdot 10^{-2}$ & 5.7279 $\cdot 10^{-2}$ & 5.5042 $\cdot 10^{-2}$ \\
$\widehat{MISE}$ & 1.1786 $\cdot 10^{-1}$ & 1.1348 $\cdot 10^{-1}$ & 9.4367 $\cdot 10^{-2}$ & 8.7674 $\cdot 10^{-2}$ & 1.3218 $\cdot 10^{-1}$ & 1.3932 $\cdot 10^{-1}$ & 1.1801 $\cdot 10^{-1}$ & 1.1574 $\cdot 10^{-1}$ \\
$\lambda$ & variable & 0.0376 & 0.0376 & 0.0376 & variable & 0.151 & 0.148 & variable \\
$\zeta^{*}$ & 0.8413 & --- & --- & --- & 0.8954 & --- & --- & --- \\
SD$(\iota=0.00)$ & 6.8077 $\cdot 10^{-2}$ & 5.7224 $\cdot 10^{-2}$ & 7.3294 $\cdot 10^{-2}$ & 5.2170 $\cdot 10^{-2}$ & 6.6888 $\cdot 10^{-2}$ & 5.8384 $\cdot 10^{-2}$ & 7.0202 $\cdot 10^{-2}$ & 3.9282 $\cdot 10^{-2}$ \\
\ \ \ \ $(\iota=0.05)$ & 8.9571 $\cdot 10^{-3}$ & 8.1636 $\cdot 10^{-3}$ & 8.9726 $\cdot 10^{-3}$ & 1.2574 $\cdot 10^{-2}$ & 8.7457 $\cdot 10^{-3}$ & 7.8246 $\cdot 10^{-3}$ & 7.6873 $\cdot 10^{-3}$ & 9.9385 $\cdot 10^{-3}$ \\
\ \ \ \ $(\iota=0.10)$ & 5.7494 $\cdot 10^{-3}$ & 4.4747 $\cdot 10^{-3}$ & 7.0503 $\cdot 10^{-3}$ & 1.0643 $\cdot 10^{-2}$ & 5.3161 $\cdot 10^{-3}$ & 3.8866 $\cdot 10^{-3}$ & 5.1410 $\cdot 10^{-3}$ & 7.6692 $\cdot 10^{-3}$ \\
\ \ \ \ $(\iota=0.15)$ & 4.8417 $\cdot 10^{-3}$ & 4.3775 $\cdot 10^{-3}$ & 6.6644 $\cdot 10^{-3}$ & 1.0569 $\cdot 10^{-2}$ & 4.4219 $\cdot 10^{-3}$ & 3.7176 $\cdot 10^{-3}$ & 4.5352 $\cdot 10^{-3}$ & 7.0767 $\cdot 10^{-3}$ \\
\hline
$n=1,000$ & & & & & & & & & \\
Bandwidth \ \ & 0.0301 & 0.0113 & 0.0234 & variable & 0.0319 & 0.0122 & 0.0300 & variable \\
${AMISE}$ & 3.19 $\cdot 10^{-2}$ & 3.4688 $\cdot 10^{-2}$ & 3.4123 $\cdot 10^{-2}$ & 3.2073 $\cdot 10^{-2}$ & 3.1536 $\cdot 10^{-2}$ & 3.1428 $\cdot 10^{-2}$ & 3.0932 $\cdot 10^{-2}$ & 2.9724 $\cdot 10^{-2}$ \\
$\widehat{MISE}$ & 8.1022 $\cdot 10^{-2}$ & 7.7727 $\cdot 10^{-2}$ & 5.6517 $\cdot 10^{-2}$ & 5.2583 $\cdot 10^{-2}$ & 9.3632 $\cdot 10^{-2}$ & 1.0096 $\cdot 10^{-1}$ & 8.1154 $\cdot 10^{-2}$ & 8.3193 $\cdot 10^{-2}$ \\
$\lambda$ & variable & 0.0376 & 0.0376 & 0.0376 & variable & 0.151 & 0.148 & variable \\
$\zeta^{*}$ & 0.8413 & --- & --- & --- & 0.8954 & --- & --- & --- \\
SD$(\iota=0.00)$ & 3.5786 $\cdot 10^{-3}$ & 3.0200 $\cdot 10^{-2}$ & 3.9697 $\cdot 10^{-2}$ & 2.7082 $\cdot 10^{-2}$ & 3.5389 $\cdot 10^{-2}$ & 3.1495 $\cdot 10^{-2}$ & 3.6970 $\cdot 10^{-2}$ & 2.0230 $\cdot 10^{-2}$ \\
\ \ \ \ $(\iota=0.05)$ & 2.7960 $\cdot 10^{-3}$ & 3.0162 $\cdot 10^{-3}$ & 3.7711 $\cdot 10^{-3}$ & 6.1781 $\cdot 10^{-3}$ & 2.6435 $\cdot 10^{-3}$ & 2.7381 $\cdot 10^{-3}$ & 2.6548 $\cdot 10^{-3}$ & 4.2529 $\cdot 10^{-3}$ \\
\ \ \ \ $(\iota=0.10)$ & 2.7418 $\cdot 10^{-3}$ & 3.0630 $\cdot 10^{-3}$ & 3.9183 $\cdot 10^{-3}$ & 6.3188 $\cdot 10^{-3}$ & 2.5042 $\cdot 10^{-3}$ & 2.5137 $\cdot 10^{-3}$ & 2.6710 $\cdot 10^{-3}$ & 4.3119 $\cdot 10^{-3}$ \\
\ \ \ \ $(\iota=0.15)$ & 2.4420 $\cdot 10^{-3}$ & 2.9077 $\cdot 10^{-3}$ & 3.9366 $\cdot 10^{-3}$ & 6.4848 $\cdot 10^{-3}$ & 2.1569 $\cdot 10^{-3}$ & 2.2396 $\cdot 10^{-3}$ & 2.4142 $\cdot 10^{-3}$ & 4.2352 $\cdot 10^{-3}$ \\
\hline
\hline
\end{tabular}
\caption[]
{Comparison of the eight estimators. We employ $m(x) = m_{k}(x) = \frac{2}{5} \left[ 3 \sin(2k\pi x) + 2 \sin(3 \pi x) \right]$, $k=3$, $f_{X}(x) = (1/0.3829)(1/\sqrt{2\pi}) \exp \left(-0.5(x-0.5)^2 \right)$, $\sigma^2(x) = (2.5 + |x - 0.5|)$ defined in $I=[0, 1]$ and Gaussian kernel.} \label{bandwidth.and.AMISE.k3}
}}
\end{center}
\end{table}
\end{landscape}
\clearpage
\thispagestyle{empty}
\clearpage
\section{Discussion and Conclusion} \label{Discussion}

In the literature, we find several studies on variance-stabilizing nonparametric estimators. Anscombe (1948) proposed a VS transformation for the bin count of a histgram. VS kernel density estimation can be achieved by applying nonparametric regression estimation to transformed bin count data, normalizing the outcome to become a probability density function, and taking the inverse of the VS transformation, as in Fan et al.(1996). Abramson (1982) proposed the global variable bandwidth in density estimation in the context of bias reduction and mentioned the possibility of a VS kernel density estimator. Fan and Gijbels (1992) also mentioned the possibility of a VS LL regression estimator using global variable bandwidth. Although stabilizing the estimator variance in itself is one definition, constructing VS nonparametric estimators enables us to create confidence bands with a constant width for an entire curve. This is also expected to stabilize the results in nonparametric estimation.

In addition to the above mentioned aspects, we proposed two methods for obtaining VS kernel regression estimators using the CC estimator. The first method employs the VS local variable bandwidth, and the other locally controls the weighting parameter of the CC estimator to make the estimator variance constant over the domain. A major difference between the two methods is that the VS bandwidth method is applicable to any kind of data, whereas the weighting method is not suitable for data whose variability in the vertical direction changes substantially over the domain. The type of kernel is an important factor for the applicability of the weighting method in variance stabilization. In general, a kernel that distributes a higher weight to more distant points can handle several types of data. We compare the performances of the two methods through simulations and find the evidence that the VS weighting method performs better than the VS bandwidth method in terms of the degree of variance stabilization when the sample size is large and the boundary effect is ignored. We also find that the strategy to minimize the estimator variance using the VS weighting method can yield AMISE values that are smaller than those obtained using the MSE-minimizng bandwidth method.

Some studies on the linear combination of nonparametric estimators can be found in addition to Choi and Hall (1998). Chen et al. (2000) applied SK methods to two-parameter locally parametric density estimators in Hojort and Jones (1996), in which a local kernel smoothed likelihood function is defined to estimate the best local parametric approximant to the true density for every $x$, and to construct the convex combination of three estimators. Because the structures of the bias and the variance brought by this method are essentially the same as those of Choi and Hall (1998), both the VS weighting and the VS local variable bandwidth methods are applicable to stabilize the estimator variance. In the context of boundary modification, Rice (1984) proposed an estimator constructed by using a linear combination of two non-skewed kernel regression estimators. One of the differences between Choi and Hall (1998) and Rice (1984) is that different bandwidth sizes are used for the components in Rice (1984), whereas a common bandwidth size is used in Choi and Hall (1998). We expect that the asymptotic variance of the estimator can be written in terms of the weighting parameter in the linear combination and the two bandwidths. We conjecture that it is possible to control one of the two bandwidths to stabilize the estimator variance and determine the weighting parameter to simultaneously make the estimator near the boundary of the same order of magnitude as that observed in the interior area.

In Section~\ref{Simulation}, we conduct simulations on the assumption that the components of the bandwidths, $f_{X}^{(i)}(x), i=1$, and $2$, $\sigma^{2}(x)$, and $m^{(j)}(x), j=2,3$, and $4$, are all true functions because our main concern is to compare the performances of the two VS methods. To obtain the data driven bandwidth $\widehat{h_{VS}^{*}}$ and weighting parameter $\widehat{\lambda_{VS}^{*}}(x)$, a plug-in method can be used; however, nonparametric estimations for $m^{(i)}(x), i=2,3$, and $4,$ are generally difficult, and the estimation process as a whole is expected to be computationally too intensive. 
\section*{Acknowledgments}
This study is financially supported by Japan Society for the Promotion of Science under a Grant-in-Aid for Young Scientists (B) JP16K17142 and a Grant-in-Aid for Scientific Research (C) 26520110.

\end{document}